\documentclass[fleqn,usenatbib]{mnras}
\usepackage[normalem]{ulem}
\usepackage{newtxtext,newtxmath}

\usepackage[T1]{fontenc}
\usepackage{ae,aecompl}

\usepackage{graphicx}	
\usepackage{amsmath}	

\usepackage{soul} 

\newcommand{\hmsun}{h^{-1}{\rm M}_\odot}
\newcommand{\hmpc}{h^{-1}{\rm Mpc}}

\newcommand{\hubunit}{\rm km~s^{-1}~Mpc^{-1}}
\newcommand{\kms}{\rm {km/s} }

\newcommand{\rrs}{R_{\rm v}^{\rm rs}}
\newcommand{\rzs}{R_{\rm v}^{\rm zs}}
\newcommand{\qrsd}{q_{\rm RSD}^s}
\newcommand{\qrsdb}{q_{\rm RSD}^l}
\newcommand{\qrsdc}{q_{\rm RSD}}
\newcommand{\qap}{q_{\rm AP}}
\newcommand{\dRc}{\delta R_{\rm v}}
\newcommand{\dR}{\delta R_{\rm v}^s}
\newcommand{\dRb}{\delta R_{\rm v}^l}

\newcommand{\q}{R_{\rm v}^{\rm zs}/R_{\rm v}^{\rm rs}}
\newcommand{\disp}{\boldsymbol{d}_{\rm v}}
\newcommand{\velv}{\boldsymbol{V}_{\rm v}}
\newcommand{\posv}{\boldsymbol{X}_{\rm v}}
\newcommand{\possvx}{X_{\rm v1}}
\newcommand{\possvy}{X_{\rm v2}}
\newcommand{\possvz}{X_{\rm v3}}
\newcommand{\posvx}{d_{\rm v1}}
\newcommand{\posvy}{d_{\rm v2}}
\newcommand{\posvz}{d_{\rm v3}}
\newcommand{\velvx}{V_{\rm v1}}
\newcommand{\velvy}{V_{\rm v2}}
\newcommand{\velvz}{V_{\rm v3}}

\newcommand{\zsim}{z}

\title[Redshift-space effects in voids I]{Redshift-space effects in voids and their impact on cosmological tests\\
Part I: the void size function}

\author[Correa et al.]{
Carlos M. Correa,$^{1,2}$\thanks{E-mail: cmcorrea@unc.edu.ar (CMC)}
Dante J. Paz,$^{1,2}$
Ariel G. S\'anchez,$^{3}$
Andr\'es N. Ruiz,$^{1,2}$
\newauthor
Nelson D. Padilla$^{4,5}$
and Ra\'ul E. Angulo$^{6,7}$
\\
$^{1}$Instituto de Astronom\'ia Te\'orica y Experimental, UNC-CONICET, Laprida 854, X5000BGR C\'ordoba, Argentina\\
$^{2}$Observatorio Astron\'omico de C\'ordoba, Universidad Nacional de C\'ordoba, Laprida 854, X5000BGR C\'ordoba, Argentina\\
$^{3}$Max-Planck-Institut f\"ur Extraterrestrische Physik, Postfach 1312, Giessenbachstr, D-85741 Garching, Germany\\
$^{4}$ Instituto de Astrof\'isica, Pontificia Universidad Cat\'olica de Chile, Av. Vicu\~na Mackenna 4860, Santiago, Chile\\
$^{5}$ Centro de Astro-ingenier\'ia, Pontificia Universidad Cat\'olica de Chile, Av. Vicu\~na Mackenna 4860, Santiago, Chile\\
$^{6}$ Donostia International Physics Centre (DIPC), Paseo Manuel de Lardizabal 4, E-20018 Donostia-San Sebastian, Spain\\
$^{7}$ IKERBASQUE, Basque Foundation for Science, E-48013 Bilbao, Spain\\
}

\date{Accepted XXX. Received YYY; in original form ZZZ}

\pubyear{2020}

\begin{document}
\label{firstpage}
\pagerange{\pageref{firstpage}--\pageref{lastpage}}
\maketitle


\begin{abstract}
Voids are promising cosmological probes.
Nevertheless, every cosmological test based on voids must necessarily employ methods to identify them in redshift space.
Therefore, redshift-space distortions (RSD) and the Alcock-Paczy\'nski effect (AP) have an impact on the void identification process itself generating distortion patterns in observations.
Using a spherical void finder, we developed a statistical and theoretical framework to describe physically the connection between the identification in real and redshift space.
We found that redshift-space voids above the shot noise level have a unique real-space counterpart spanning the same region of space, they are systematically bigger and their centres are preferentially shifted along the line of sight.
The \textit{expansion effect} is a by-product of RSD induced by tracer dynamics at scales around the void radius, whereas the \textit{off-centring effect} constitutes a different class of RSD induced at larger scales by the global dynamics of the whole region containing the void.
The volume of voids is also altered by the fiducial cosmology assumed to measure distances, this is the \textit{AP change of volume}.
These three systematics have an impact on cosmological statistics.
In this work, we focus on the void size function.
We developed a theoretical framework to model these effects and tested it with a numerical simulation, recovering the statistical properties of the abundance of voids in real space.
This description depends strongly on cosmology.
Hence, we lay the foundations for improvements in current models of the abundance of voids in order to obtain unbiased cosmological constraints from redshift surveys.
\end{abstract}







\section{Introduction}
\label{sec:intro}

Cosmic voids are vast underdense regions.
Since they are the largest observable structures, voids are proving to be powerful cosmological laboratories, as they encode information about the expansion history and geometry of the Universe.
This is very valuable, since one of the major goals of modern cosmology is to understand the nature of the accelerated expansion of space, which is believed to be caused by a dark energy component.

Voids offer two special advantages over the high density regime, which make them attractive to theoretical modelling and observational test designing.
On the one hand, void dynamics can be treated linearly, assuming spherical symmetry for the velocity and density fields.
In this way, it is easier to model systematics such as redshift-space distortions \citep{clues2,aprsd_hamaus1,rsd_cai,aprsd_hamaus2,rsd_achitouv1,rsd_achitouv2,rsd_chuang,rsd_hamaus,rsd_hawken1,rsd_achitouv3,aprsd_correa,rsd_nadathur,aprsd_nadathur,reconstruction_nadathur,rsd_hawken2}. 
On the other hand, modified gravity theories predict deviations from general relativity to be most pronounced in these unscreened low density environments
(\citealt{ellipticity_bos}; \citealt{modgr_li}; \citealt{modgr_clampitt}; \citealt{modgr_clifton}; \citealt{modgr_barreira}; \citealt{modgr_cai}; \citealt{modgr_lam}; \citealt{modgr_zivick}; \citealt{modgr_achitouv}; \citealt{modgr_joyce}; \citealt{modgr_koyama}; \citealt{modgr_cautun}; \citealt{modgr_falck}; \citealt{modgr_sahlen}; \citealt{modgr_davies}; \citealt{modgr_paillas}).
These theories offer alternative explanations for the dark energy problem.

Regarding the exploitation of voids as cosmological probes, there are two primary statistics: the void size function and the void-galaxy correlation function.
The void size function \citep[hereafter VSF]{SvdW,abundance_furlanetto,jennings_Vdn,abundance_achitouv,abundance_pisani,abundance_ronconi1,abundance_bias_contarini,abundance_ronconi2,abundance_verza}, describes the abundance of voids, and can be modelled using the excursion set formalism combined with the spherical expansion of matter underdensities derived from perturbation theory.
The void-galaxy correlation function (hereafter VGCF), on the other hand, characterises the void density field when considered at small to intermediate scales (i.e. the one void term).
For the moment, there is not a first-principle model for it.
There are, however, many parametric and empirical ones in the literature (e.g. \citealt{clues2,density_hamaus,density_nadathur,aprsd_correa}).

Both, the VSF and the VGCF, are primarily affected by two redshift-space systematics.
The first is a dynamical effect, the \textit{redshift-space distortions} (\citealt{kaiser_rsd}, hereafter RSD).
This effect is
caused by the peculiar velocities of matter tracers, which manifest in observations as an additional and indistinguishable contribution to the measured redshifts, which in turn induce a distorted estimation of the line-of-sight (hereafter LOS) coordinates.
The second is a geometrical effect, the \textit{Alcock-Paczy\'nski effect} (\citealt{ap}, hereafter AP), caused by the selection of a fiducial cosmology when transforming observables from a catalogue (angles and redshifts) into a Mpc-scale. 
These systematics can be physically modelled and are cosmology dependent.
Therefore, they are a source of valuable information when designing cosmological tests.
For instance, RSD in the VGCF can be modelled with the Gaussian streaming model \citep{clues2,rsd_cai,rsd_achitouv1,rsd_achitouv2,rsd_chuang,rsd_hamaus,rsd_hawken1,rsd_achitouv3,rsd_nadathur,reconstruction_nadathur,rsd_hawken2}. 
For a full modelling taking into account both RSD and the AP effect, see \cite{aprsd_hamaus1,aprsd_hamaus2,aprsd_correa} and \cite{aprsd_nadathur}.
In \cite{aprsd_correa}, in addition, we explain a third important systematic: the \textit{mixing of scales} due to the binning scheme used to measure the VGCF.

There are different classes of void finders (see the seminal paper of \citealt{voidID_colberg} for a thorough comparison of different methods).
In our case, we use the so called spherical void finder \citep{voids_padilla}, based on the integrated density contrast of underdense regions assuming spherical symmetry (we use in this work the modified version of \citealt{clues3}).
As examples of different techniques, see \citealt{zobov} for the commonly used \textsc{ZOBOV} void finder based on the watershed algorithm, and \citealt{voidID_elyiv} for a method based on the dynamical properties of tracers.
Despite the intrinsic differences between the available methods, since in general they are based on the density and dynamics of matter tracers, it is expected that RSD and the AP effect also affect void properties.
This was specially noted in \cite{aprsd_nadathur} and \cite{reconstruction_nadathur}, where they show that there are four hypotheses commonly assumed when modelling RSD, which are only valid for voids identified in real space, and are violated for voids identified in redshift space.
Specifically, these hypotheses are: (1) conservation of void number, (2) isotropy of the density field, (3) isotropy of the velocity field, and (4) invariability of void centre positions.
In \cite{reconstruction_nadathur}, they explain that this problem can be tackled in two different ways: (i) using a reconstruction technique \citep{reconstruction_eisenstein}, or (ii) analysing physically the void finding mechanism.
They focused on the first approach.
Reconstruction is an algorithm to recover the real-space positions of galaxies from redshift space based on the Zel'dovich approximation.
This technique has been used for BAOs analyses as well.
The idea is to apply the reconstruction before performing the void finding step.
As this is a cosmology dependent procedure, it can be used to measure the growth rate factor if the reconstruction plus the void finding step are applied iteratively.

In this paper, instead, we focus on the second approach, namely, we analyse the void finding method in order to understand physically the redshift-space systematics affecting voids.
To do this, we use a spherical void finder and make a statistical comparison between the resulting real and redshift-space voids.
We find two relevant results.
First, there is a one-to-one relationship between voids.
This means that each redshift-space void has a unique real-space counterpart and vice versa, spanning the same region of space.
In this sense, condition (1) of void number conservation is not violated.
This is a consequence of our void definition: voids are large spherical regions with very low integrated density, and hence, mostly expanding.
Second, redshift-space voids are systematically bigger than their real-space counterparts, and their centre positions are shifted preferentially along the LOS.
These phenomenons can be attributed to two physical effects: \textit{expansion} and \textit{off-centring}, which in turn, can be theoretically described based on both tracer and void dynamics.
Moreover, this description depends strongly on cosmology.

This paper is the first of two publications concerning the impact of redshift-space effects in voids on cosmological statistics.
Here, we study the void size function.
We leave the implications of these effects on the void-galaxy correlation function for the second part.
Up to our knowledge, this is the first time that redshift-space systematics on the VSF are treated.
The community has concentrated their efforts on modelling the true underlying real-space VSF with the excursion set formalism.
The intention of this work is to lay the foundations for a full modelling, leaving for future investigation to link both developments.
In view of the new generation of spectroscopic surveys, such as the Baryon Oscillation Spectroscopic Survey \citep[BOSS]{boss}, and the future Dark Energy Spectroscopic Instrument Survey \citep[DESI]{desi2} and Euclid \citep{euclid}, which will probe our Universe covering a volume and a redshift range without precedents, it is extremely important to detect and model all the z-space systematics that affect the spatial distribution of galaxies in order to obtain unbiased cosmological constraints.
In this sense, the z-space effects studied in this work must be incorporated in any analysis of RSD around voids in order to successfully exploit these cosmic structures as cosmological probes.
Furthermore, besides its cosmological importance, these z-space systematics encode key information about the structural and dynamical nature of voids.

This paper is organised as follows.
In Section~\ref{sec:data}, we describe the data set, that is, the numerical N-body simulation and the void catalogues.
In Section~\ref{sec:map}, we explain the bijective mapping between real and redshift-space voids.
In Section~\ref{sec:effects}, we explain theoretically the expansion and off-centring effects, along with the additional AP change of volume.
Then, in Section~\ref{sec:stat}, we provide a statistical analysis confirming them.
After that, in Section~\ref{sec:vsf}, we analyse the implications of these effects on the VSF as a cosmological test.
Finally, we summarize and discuss our results in Section~\ref{sec:conclusions}.


\section{Data set}
\label{sec:data}


\subsection{Simulation setup}
\label{subsec:data_sim}

We used the Millennium XXL N-body simulation \citep[hereafter MXXL]{mxxl} which extends the previous Millennium and Millennium-II simulations \citep{millennium,millennium2} and follows the evolution of $6720^3$ dark matter particles inside a cubic box of length $3000~\hmpc$.
The particle mass is $8.46\times10^9~\hmsun$ in a flat $\Lambda$CDM cosmology with the same cosmological parameters as the previous runs: $\Omega_m=0.25$, $\Omega_\Lambda=0.75$, $\Omega_b=0.045$, $\Omega_\nu=0.0$, $h=0.73$\footnote{The Hubble constant is parametrised as $H_0=100~h~\hubunit$. All distances and masses are expressed in units of $\hmpc$ and $\hmsun$ respectively.}, $n_s=1.0$ and $\sigma_8=0.9$.
We used the snapshot belonging to redshift $\zsim=0.99$, which will be assumed as the mean redshift of the sample.

Dark matter haloes were chosen as tracers, which were identified as groups of more than $60$ particles using a friends-of-friends algorithm with a linking length parameter of $0.2$ times the mean inter particle separation.
We selected a lower mass cut of $5\times10^{11}~\hmsun$, finding in this way $136,688,808$ haloes.

Positions $\boldsymbol{x} = (x_1,x_2,x_3)~[\hmpc]$ and peculiar velocities $\boldsymbol{v} = (v_1,v_2,v_3)~[\kms]$ of haloes in real space were available to quantify the effects of distortions.
In order to generate RSD, we treated the $x_3$-axis of the simulation box as the LOS direction, assuming the distant observer approximation, where changes in the LOS direction with the observed angles on the plane of the sky (hereafter POS) can be neglected.
For the purposes of this work, this is a fair assumption for two main reasons.
On the one hand, the redshift of the snapshot is far enough considering that we only analyse void-centric distances less than $40~\hmpc$.
On the other hand, although the box length is large, we treat it as the simplest mock, with no intention to treat observational systematics nor light cone effects.
The volume of the simulation simply allows to have a large sample of voids to detect the different redshift-space systematics.
We applied the following equation to shift the LOS coordinates of haloes from real to redshift space:
\begin{equation}
    \Tilde{x}_3 = x_3 + v_3 \frac{(1 + \zsim)}{H(\zsim)},
	\label{eq:halo_zspace}
\end{equation}
where $\Tilde{x}_3$ denotes the shifted $x_3$-coordinate, and $H(z)$ is the Hubble parameter, which for a flat-$\Lambda$CDM cosmology can be expressed in terms of the cosmological parameters as follows:
\begin{equation}
    H(z) = 100~h~\sqrt{\Omega_m(1+z)^3 + \Omega_\Lambda}.
    \label{eq:hubble}
\end{equation}
For our flat case,
\begin{equation}
    \Omega_\Lambda = 1 - \Omega_m.
    \label{eq:flat}
\end{equation}


\subsection{Void catalogues}
\label{subsec:data_voids}

We applied the spherical void finder described in \citet{clues3}, which is a modified version of the algorithm of \citet{voids_padilla}.
Since the goal of this paper is to understand the implications of redshift-space systematics on void identification, we detail below the steps of this procedure:

\begin{enumerate}
    \item
    A Voronoi tessellation is performed to obtain an estimation of the density field: each halo has an associated cell with volume $V_{\rm cell}$, and a density given by the inverse of that volume: $\rho_{\rm cell} = 1/V_{\rm cell}$.
    We used a parallel version of the public library \textsc{voro++} \citep{voropp}.
    \\
    \item
    A first selection of underdense regions is done by selecting all Voronoi cells which satisfy the criterion $\delta_{\rm cell} := \rho_{\rm cell}/\bar{n} - 1 < -0.7$, where $\bar{n}$ is the mean number density of haloes.
    Each underdense cell is considered the centre of a potential void.
    \\
    \item
    Centred on each candidate, the integrated density contrast $\Delta(r) = \delta(<r)$ is computed in spheres of increasing radius $r$ until the overall density contrast satisfies a redshift dependent threshold of $\Delta_{\rm id} = -0.853$ for $\zsim=0.99$, obtained from the spherical collapse model \citep{sphcollapse1,sphcollapse2} by fixing a final spherical perturbation of $\Delta_{\rm id} = -0.9$ for $z=0$.
    The choice of $\Delta_{\rm id}$ is motivated by previous studies that use voids identified from biased tracers, namely, haloes or galaxies.
    For dark matter voids \citep{SvdW,jennings_Vdn}, there is a theoretical threshold: $\Delta_{\rm id} = -0.8$, that corresponds to the moment of shell-crossing in the expansion process, which is taken as the moment of void formation.
    However, it is not trivial to extrapolate this value for the case of voids identified from biased tracers.
    Some studies provide a firm evidence about a linear bias relation between dark matter and tracer voids \citep{abundance_furlanetto,bias_pollina1,bias_pollina2,bias_pollina3,abundance_bias_contarini}.
    As pointed out by \cite{abundance_bias_contarini}, assuming that dark matter and tracer voids have the same radii when the phenomenon of shell-crossing occurs, implies that the latter have a lower embedded density contrast.
    Therefore, if the bias is greater than $1$, then the density contrast can reach values so low that the phenomenon of shell-crossing might not even happen.
    Hence, it is a common practice to define voids as empty as possible.
    Previous works using the spherical void finder have demonstrated that choosing $\Delta_{\rm id} = -0.9$ for $z = 0$ (and extrapolating this value to the corresponding redshift according to the spherical collapse model), give a sample of voids with a well characterised dynamics and suitable for cosmological studies \citep{clues1, clues2, aprsd_correa}.
    \\
    \item
    Once these first void candidates are identified, step (iii) is repeated, but starting in a randomly displaced centre proportional to $0.25$ times the radius of the candidate.
    Then, the void centre is updated to a new position if the new radius is larger than the previous one.
    This process is repeated iteratively until convergence to a sphere with maximum radius is achieved.
    We adopted the criterion that the optimal sphere is obtained if the algorithm can not find a bigger one during a lapse of $50$ iterations.
    This normally takes between $200$ and $300$ iterations in total.
    In this way, this procedure mimics a random walk around the original centre in order to obtain the largest possible sphere in that local minimum of the density field.
    \\
    \item
    Finally, the list of void candidates is cleaned so that each resulting sphere does not overlap with any other.
    This cleaning is done by ordering the list of candidates by size and starting from the largest one.
    The final result is a catalogue of non-overlapping spherical voids with radii $R_{\rm v}$ and overall density contrast $\Delta(R_{\rm v}) = \Delta_{\rm id}$.
\end{enumerate}

We applied the void finder in two ways, resulting in two types of catalogues.
In the first case, we adopted the same cosmology of the MXXL simulation in order to compute distances and densities, needed in void definition.
In turn, we performed the identification both in real and redshift space (hereafter r-space and z-space respectively), in order to study the impact of RSD.
We will refer to these catalogues as the \textit{true-cosmology (TC) void catalogues}.

In the second case, we modified the simulation coordinate system according to two different cosmologies, in order to study the impact of the combined RSD and AP effects.
Specifically, we fixed all the MXXL global parameters with the exception of $\Omega_m$ and $\Omega_\Lambda$, for which a lower and an upper fiducial values for $\Omega_m$ were chosen: $\Omega_m^l = 0.20$ and $\Omega_m^u = 0.30$, in such a way that the cosmology was still flat, i.e., Eq.~(\ref{eq:flat}) was still valid.
In this case, the identification was only performed in z-space (see Section~\ref{sec:vsf} for more details).
We will refer to these catalogues as the \textit{fiducial-cosmology (FC) void catalogues}.
Table~\ref{tab:catalogues} shows the main characteristics of the void catalogues and selected samples used in this work.

\begin{table*}
\centering
\caption{
Main characteristics of the void samples used in this work.
From left to right: sample name, catalogue type, cosmology used in void definition, $\Omega_m$ choice, space where the identification was performed, completeness regarding the bijective filtering, number of voids, and type of systematics taken into account.
}
\label{tab:catalogues}
\begin{tabular}{cccccccc}
\hline
Sample & Catalogue & Cosmology & $\Omega_m$ & Space & Completeness & Number of voids & Systematics \\
\hline
\hline
TC-rs-f & TC & MXXL     & 0.25 & r-space & full      & 463,690 & none \\ 
TC-rs-b & TC & MXXL     & 0.25 & r-space & bijective & 318,784 & none \\
TC-zs-f & TC & MXXL     & 0.25 & z-space & full      & 455,482 & RSD \\
TC-zs-b & TC & MXXL     & 0.25 & z-space & bijective & 318,784 & RSD \\
FC-l    & FC & Fiducial & 0.20 & z-space & full      & 375,560 & AP + RSD \\ 
FC-u    & FC & Fiducial & 0.30 & z-space & full      & 526,552 & AP + RSD \\
\hline
\end{tabular}
\end{table*}


\section{Bijective mapping}
\label{sec:map}

\begin{figure*}
    \includegraphics[width=\columnwidth]{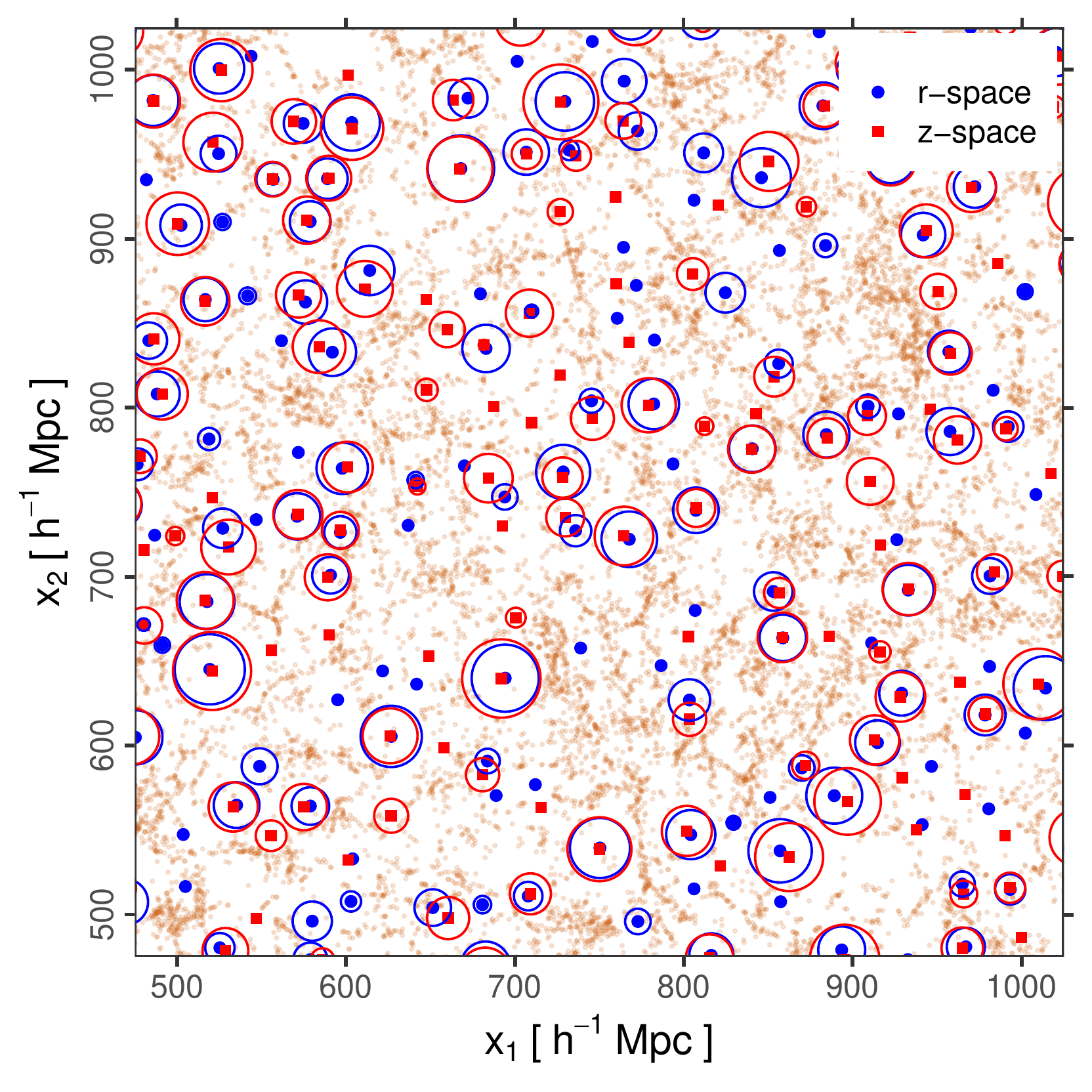}
    \includegraphics[width=\columnwidth]{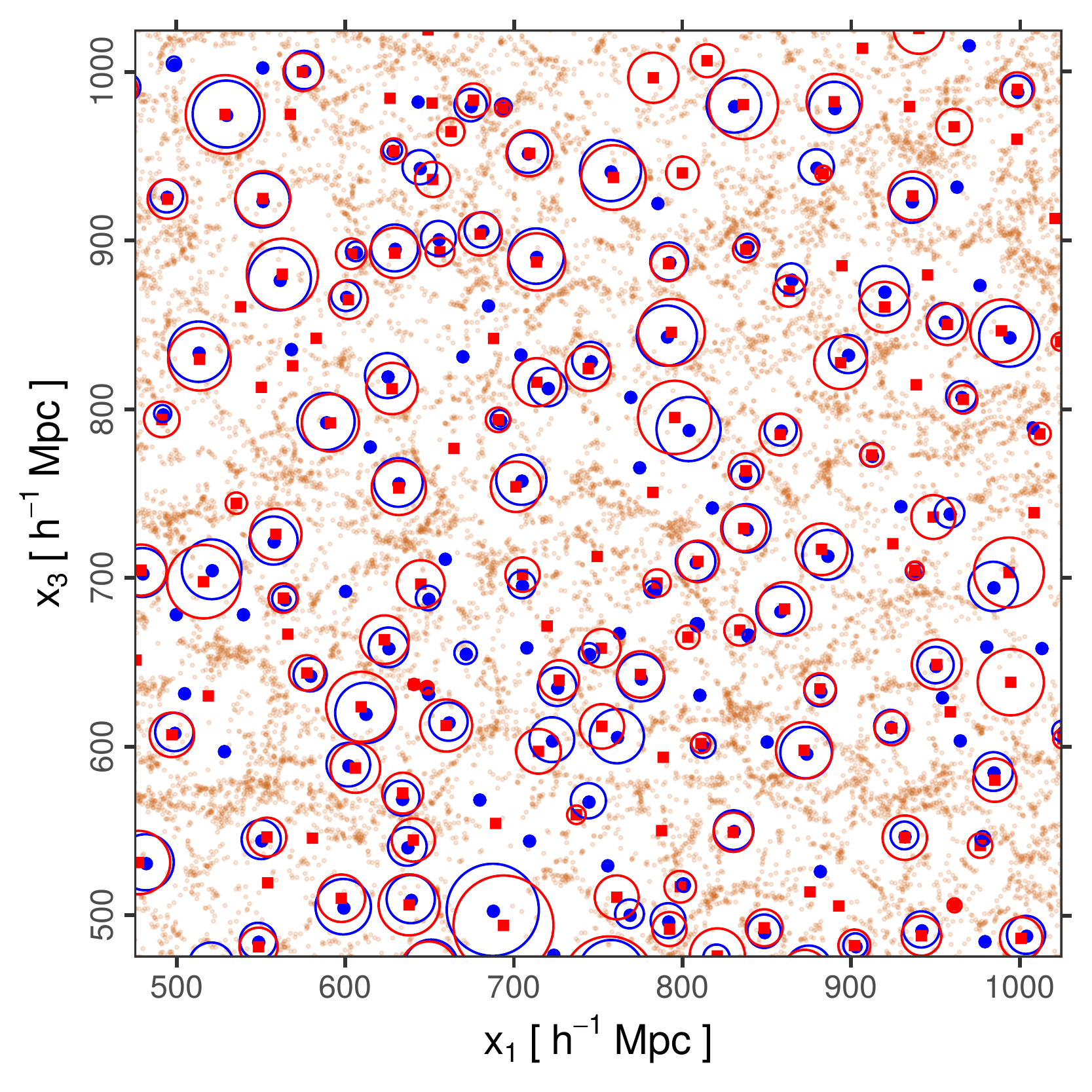}
    \caption{
    Slices of the Millennium XXL simulation box showing the distribution of haloes and voids from the TC samples of Table~\ref{tab:catalogues}.
    Real-space void centres are represented with blue dots, whereas redshift-space centres, with red squares.
    Bijective voids, in turn, are represented with circles, the intersections of the spherical voids with the mid-plane of the slice.
    \textit{Left panel.}
    Slice in the range $500 \leq x_1/\hmpc \leq 1000$, $500 \leq x_2/\hmpc \leq 1000$ and $95 \leq x_3/\hmpc \leq 105$, a representation of the plane-of-sky distribution.
    \textit{Right panel.}
    Slice in the range $500 \leq x_1/\hmpc \leq 1000$, $95 \leq x_2/\hmpc \leq 105$ and $500 \leq x_3/\hmpc \leq 1000$, a representation of the line-of-sight distribution.
    Bijective voids span the same regions in both spaces.
    Voids expand and their centres shift when they are mapped from r-space into z-space.
    }
    \label{fig:mxxl_slice}
\end{figure*}

We begin the analysis with a visual inspection of r-space and z-space voids.
Fig.~\ref{fig:mxxl_slice} shows two slices of the MXXL simulation box using the TC void catalogues.
The aim of this section is to study the impact of RSD alone, postponing the analysis of the combined RSD and AP effects until Section~\ref{sec:vsf}, where we will use the FC void catalogues.
The left panel is a slice in the range $500 \leq x_1/\hmpc \leq 1000$, $500 \leq x_2/\hmpc \leq 1000$ and $95 \leq x_3/\hmpc \leq 105$.
Hence, it is a representation of the POS distribution of haloes and voids.
Here, r-space void centres are represented with blue dots, whereas z-space centres, with red squares.
The right panel, on the other hand, is a slice in the range $500 \leq x_1/\hmpc \leq 1000$, $95 \leq x_2/\hmpc \leq 105$ and $500 \leq x_3/\hmpc \leq 1000$, i.e. it shows the LOS distribution of haloes and voids.
From the figure, it is clear that z-space and r-space voids approximately span the same regions of space.

In order to link r-space and z-space voids, we looked for their correspondence by cross-correlating both catalogues.
Specifically, for each z-space centre, we picked the nearest r-space centre with the condition that it must lay inside $1~\rzs$ ($\rrs$ and $\rzs$ will denote r-space and z-space radii respectively).
Then, we filtered those voids if no partner could be found.
Note that this z-space $\longrightarrow$ r-space mapping is a well defined function, since the condition of the nearest r-space neighbour assigns only one object to each z-space void.
Furthermore, this mapping is also injective, since the non-overlapping condition implies that each r-space void can only be reached by a single z-space one.
Even further, the filtering condition guarantees then a one-to-one relationship between z-space and r-space voids.
For this reason, these voids constitute what we call the \textit{bijective samples} (TC-rs-b and TC-zs-b in Table~\ref{tab:catalogues}).
Note that, by construction, these samples have the same number of elements.
Moreover, it is ensured in this way that a void and its associated counterpart span the same region of space.
In order to distinguish the bijective samples from the entire catalogues, we will refer to the last ones as the \textit{full samples} (TC-rs-f and TC-zs-f in Table~\ref{tab:catalogues}).
Going back to Fig.~\ref{fig:mxxl_slice}, bijective voids are represented with circles around their centres, which correspond to the intersections of the spherical voids with the mid-plane of the slice.
The rest are voids of the full samples without a partner in the other space.

\begin{figure*}
    \includegraphics[width=\columnwidth]{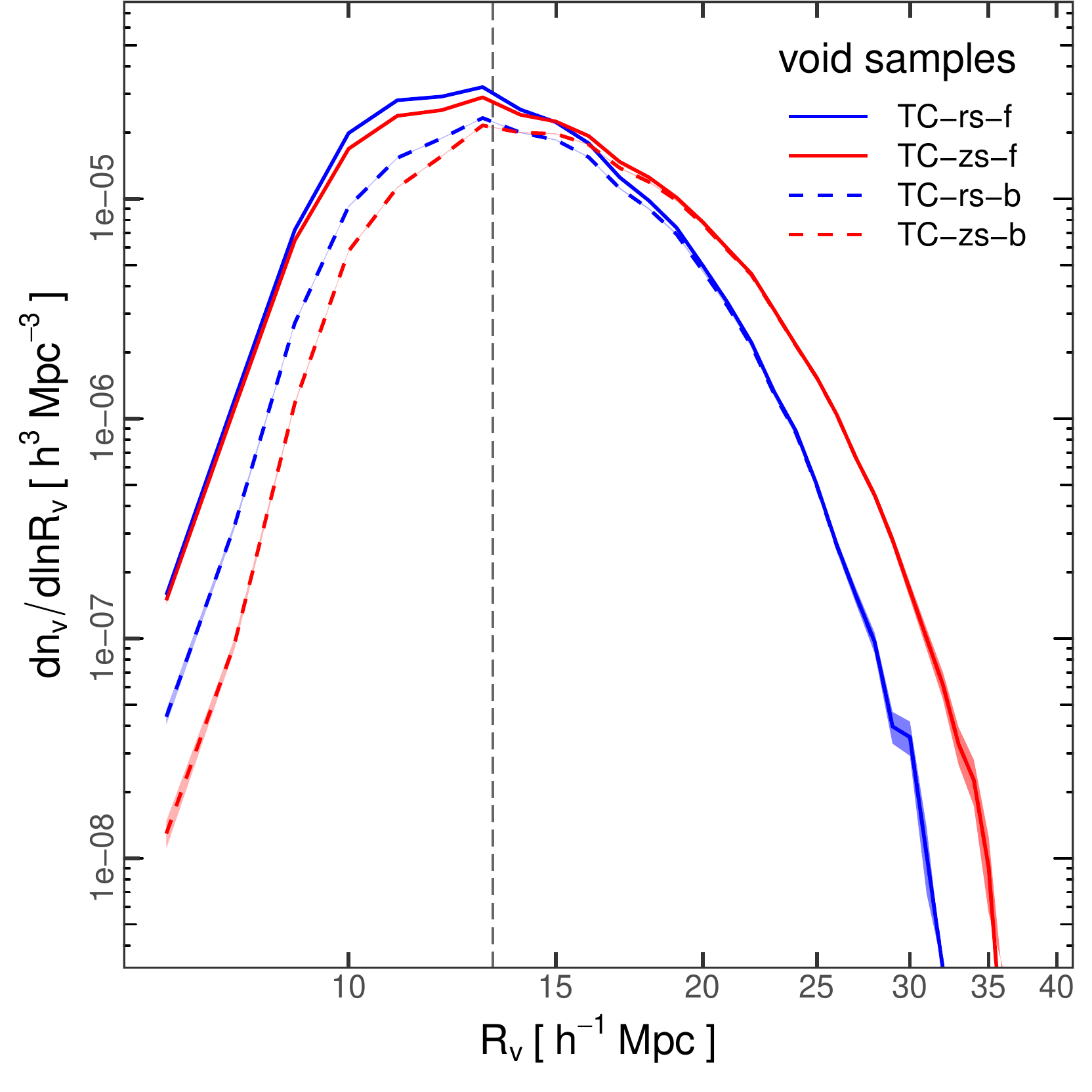}
    \includegraphics[width=\columnwidth]{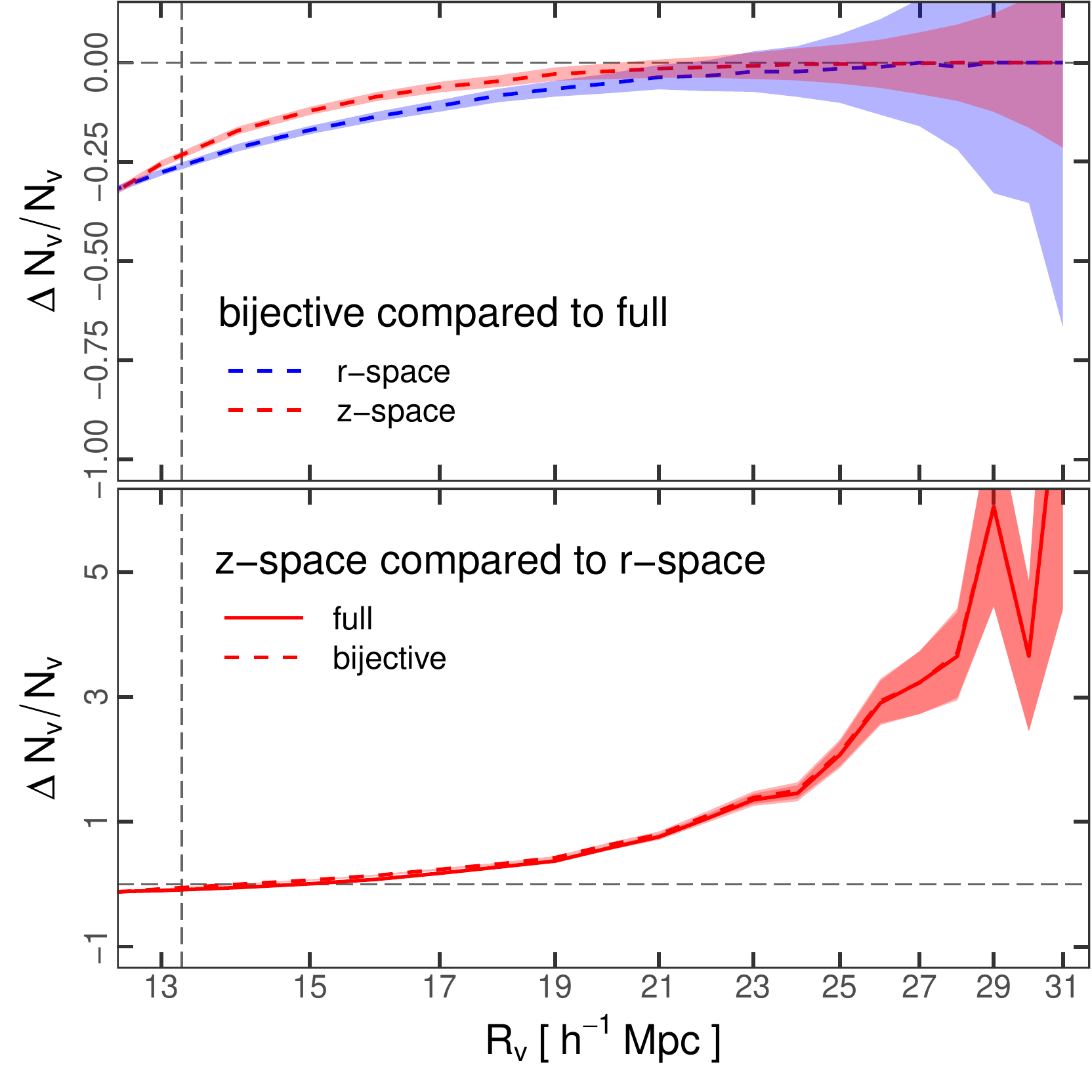}
    \caption{\textit{Left panel.} Void size functions of the TC void samples.
    The solid lines represent the VSFs of the full samples, both in r-space (blue) and z-space (red).
    The dashed lines, the VSFs of the bijective samples.
    The vertical dashed line is the median of the z-space full sample (TC-zs-f), which separates the small voids dominated by shot noise (on the left) from the large ones relevant for cosmological analyses (on the right).
    \textit{Right-upper panel.} Fractional differences of void counts between the bijective and full samples.
    It quantifies the void loss in the z-space $\longrightarrow$ r-space mapping.
    Note that large voids are almost bijective.
    \textit{Right-lower panel.} Fractional differences of void counts between the z-space and r-space samples.
    }
    \label{fig:TC_VSF}
\end{figure*}

In order to enquire deeper into the relation between r-space and z-space voids, the left panel of Fig.~\ref{fig:TC_VSF} shows the void size functions of the four TC void samples (r-space and z-space, with their full and bijective versions).
The VSFs were computed from the radius distribution of each sample, expressing the void counts as comoving differential number densities, $dn_{\rm v}$, and normalising them by the logarithmic sizes of the radius bins, $d\mathrm{ln}R_{\rm v}$.
The solid lines represent the abundances of the full samples, both in r-space (blue) and z-space (red), whereas the dashed lines, the abundances of the bijective samples.
In all cases, the error bands were calculated from Poisson errors in the void counting process.
The vertical dashed line represents the median of the z-space full sample, and will serve as a reference line throughout the work.
This value is equal to $13.26~\hmpc$ ($2.28$ in units of the mean interparticle separation.)
The qualitative behaviour of the VSFs is consistent with previous studies, where we can distinguish two main behaviours separated by the vertical line: (i) on the left, small voids dominated by shot noise, and (ii) on the right, large voids decreasing their number as the radius increases with a functional shape similar to those predicted by theory (see \citealt{SvdW} and \citealt{jennings_Vdn}).
Small voids, in this sense, are not reliable for any cosmological analysis, hence, we will mainly focus on large voids throughout this work.

Note that the full and bijective abundances (solid and dashed lines) tend to the same values as the radius increases.
This means that the loss of voids in the z-space $\longrightarrow$ r-space mapping is only significant in the region dominated by shot noise, whereas all large and relevant voids are conserved.
This is better shown in the right-upper panel of Fig.~\ref{fig:TC_VSF}, where we present the corresponding fractional differences of void counts between the full and bijective samples:
$\Delta N_{\rm v}/N_{\rm v} = (N_{\rm v}^{\rm bij} - N_{\rm v}^{\rm full})/N_{\rm v}^{\rm full}$.
For all radii of interest, the loss of voids decreases as the radius increases, being less than
$25\% ~ (\Delta N_{\rm v}/N_{\rm v} < 0.25)$
in the worst case.
We arrive here at the first and one of the most important conclusions of this work: large voids identified in an observational catalogue are true voids, i.e., they have a real-space counterpart.
Therefore, it is valid to treat the full and bijective samples indistinctly in their properties.

Comparing now the r-space and z-space abundances (blue and red lines), it is clear that they are very different.
In particular, the corresponding fractional differences $\Delta N_{\rm v}/N_{\rm v} = (N_{\rm v}^{\rm zs} - N_{\rm v}^{\rm rs})/N_{\rm v}^{\rm rs}$ (right-lower panel) increase as the radius increases, and can be very high for the largest sizes.
However, in the context of the bijective mapping, this means that these differences can only be attributed to some physical effect that voids suffer when they are mapped from r-space into z-space.
In particular, note that for each radius, z-space voids are systematically bigger than their r-space counterparts.
This hints of an expansion effect.


\section{Theoretical framework}
\label{sec:effects}

The goal of this section is to study theoretically the possible physical mechanisms responsible of the transformation of r-space voids into their associated z-space counterparts.
We will do this in the context of the four hypotheses commonly assumed to model RSD around voids noticed by \cite{reconstruction_nadathur}, which are only valid for voids identified in real-space, but are violated for voids identified in redshift-space:
\begin{enumerate}
    \item[(1)] void number conservation,
    \item[(2)] isotropy of the density field,
    \item[(3)] isotropy of the velocity field,
    \item[(4)] invariability of centre positions.
\end{enumerate}
In next section, we will provide statistical evidence of the framework developed here.


\subsection{Void number conservation}
\label{subsec:effects_conservation}

Strictly speaking, condition (1) of void number conservation is violated.
This is evident for the r-space and z-space full samples, since they have different number of voids (see Table~\ref{tab:catalogues}) and different void abundances (Fig.~\ref{fig:TC_VSF}).
Nevertheless, condition (1) is not violated in the context of the bijective mapping that we defined.
This is supported by two reasons.
First, bijective voids are, by definition, the same entities spanning the same regions of space.
This is why the bijective samples have the same number of voids.
Second, large and relevant voids are conserved after this mapping.


\subsection{Expansion effect}
\label{subsec:effects_expansion}

\begin{figure*}
	\includegraphics[width=55mm]{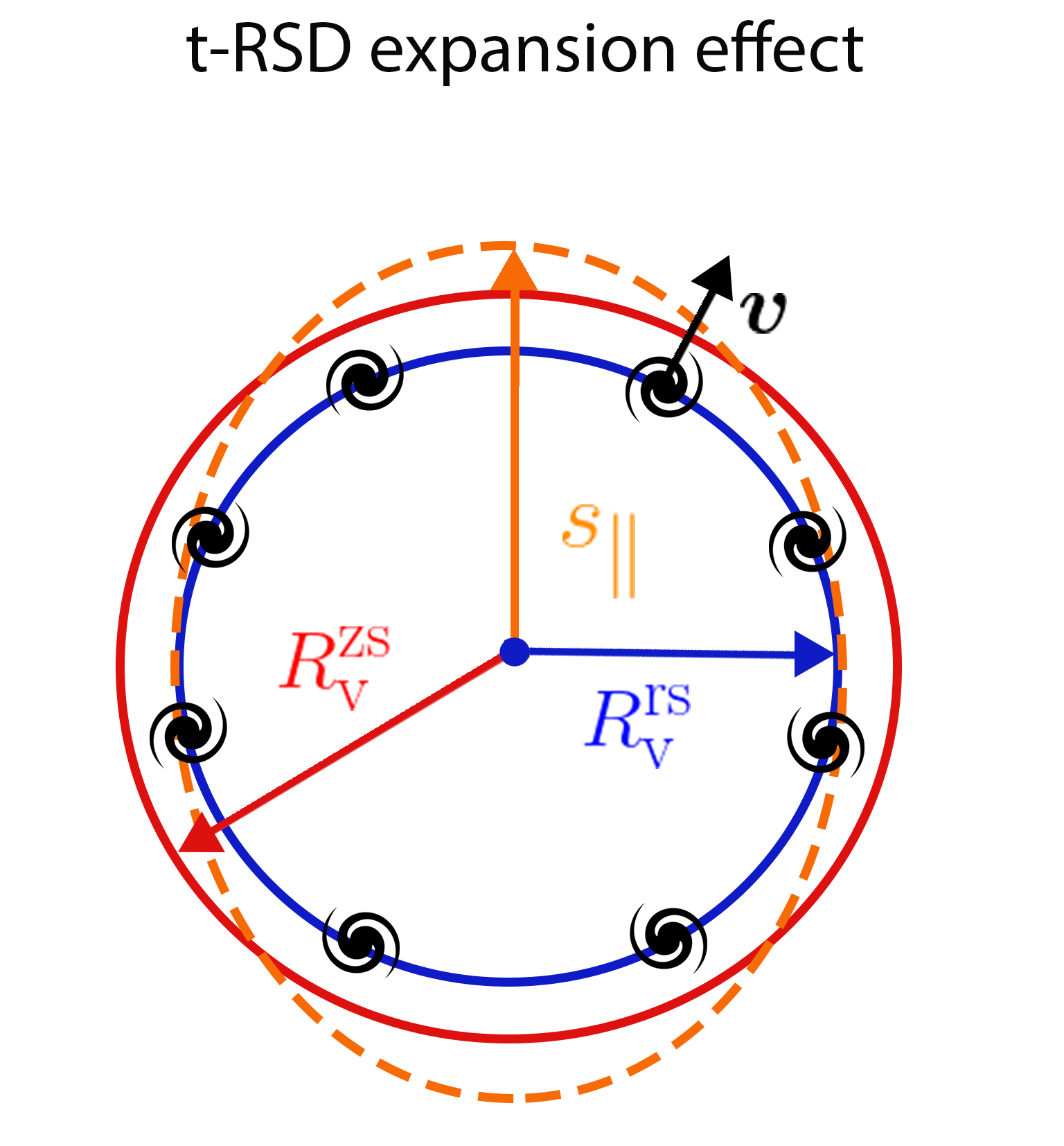}
	\includegraphics[width=55mm]{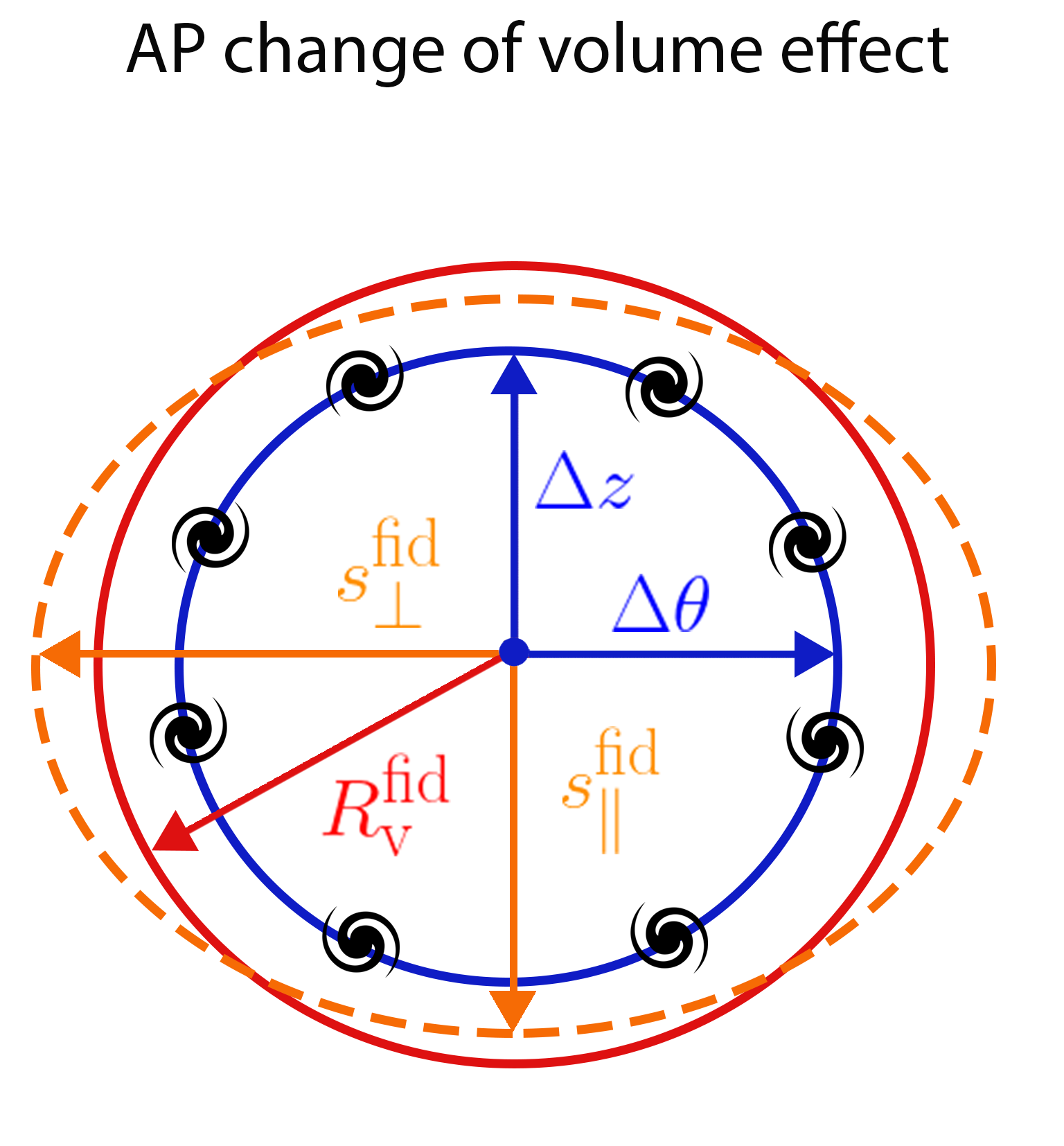}
	\includegraphics[width=55mm]{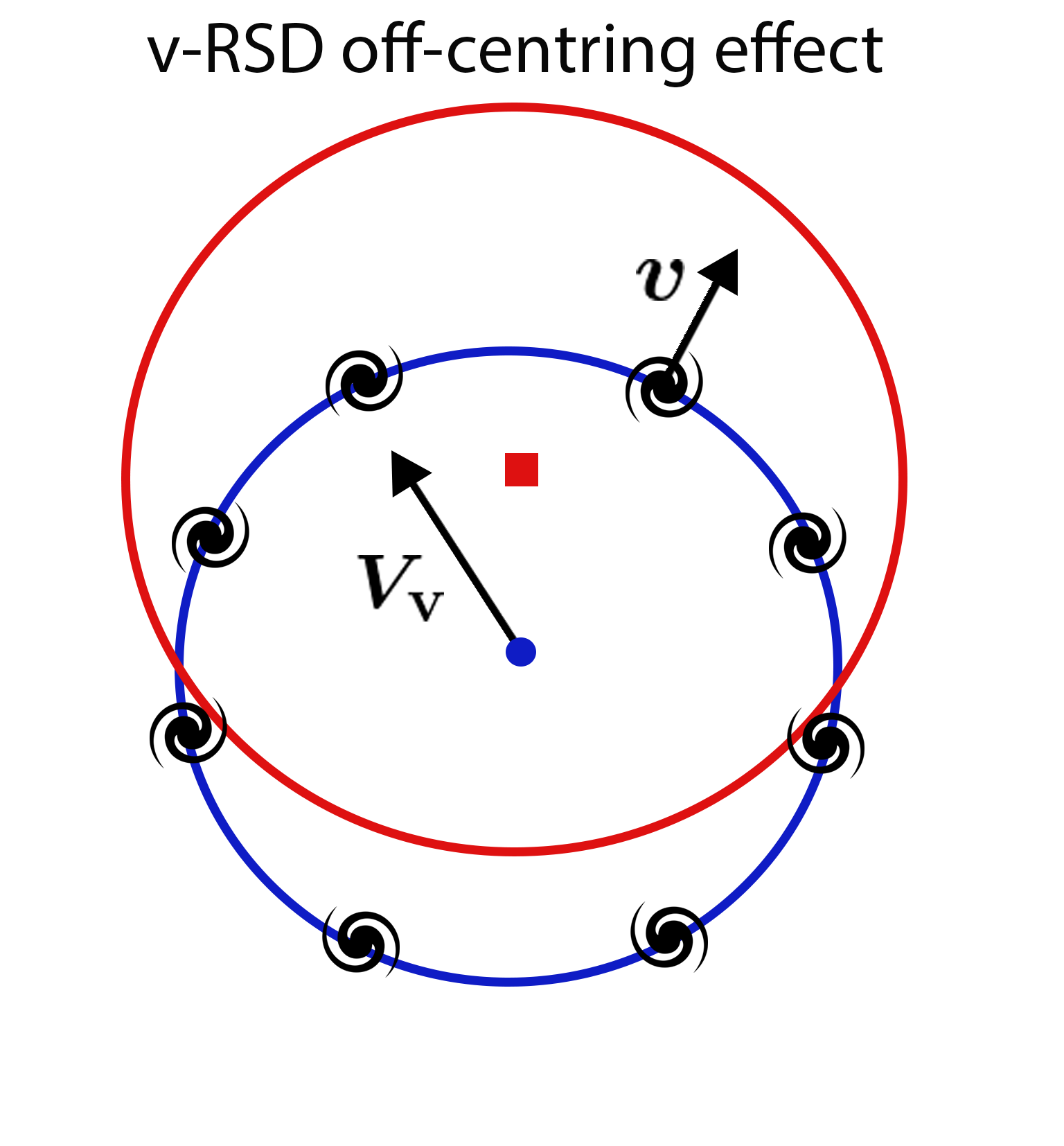}
    \caption{
    Schematic illustration of the three z-space effects in voids.
    In all panels, a hypothetical spherical r-space void of radius $\rrs$ is represented with a blue solid circle with some galaxies.
    The LOS direction is vertical.
    \textit{Left panel. Expansion effect.}
    The r-space void appears elongated along the LOS in z-space with LOS semi-axis $s_\parallel$ (orange dashed ellipse).
    The void finder identifies an equivalent sphere of radius $\rzs$ (red solid circle).
    This effect is a by-product of tracer dynamics at scales around the void radius (t-RSD).
    \textit{Middle panel. AP change of volume.}
    The r-space void is distorted into an ellipsoid with LOS semi-axis $s_\parallel^{\rm fid}$ and POS semi-axis $s_\perp^{\rm fid}$ (orange dashed ellipse).
    The void finder identifies an equivalent sphere of radius $R_{\rm v}^{\rm fid}$ (red solid circle).
    This is a geometrical effect caused by the selection of a fiducial cosmology in order to transform the observable dimensions of the void, an angular radius $\Delta \theta$ and a redshift radius $\Delta z$, into a Mpc-scale (no t-RSD are included here for a better comprehension).
    \textit{Right panel. Off-centring effect.}
    The r-space void centre appears shifted along the LOS in z-space.
    This effect is a by-product of void dynamics (v-RSD), i.e. the global dynamics of the whole region containing the void.
    The three effects can be treated independently.
    }
    \label{fig:effects}
\end{figure*}

In Section~\ref{sec:map} we showed that z-space voids are systematically bigger than their r-space counterparts.
This suggests that voids expand when they are mapped from r-space into z-space.
The left panel of Fig.~\ref{fig:effects} depicts schematically this \textit{expansion effect}.
A spherical r-space void of radius $\rrs$ (represented with a blue solid circle with some galaxies), appears elongated along the LOS in z-space due to the RSD induced by the LOS components of the peculiar velocities of the tracers surrounding it.
The r-space sphere has been transformed into a z-space ellipsoid (orange dashed ellipse in the figure) with semi-axes $(s_\perp, s_\parallel)$, where $s_\perp$ is the POS semi-axis (equal for both $x_1$ and $x_2$ directions), and $s_\parallel$, the LOS semi-axis.

We derive now analytical expressions for the semi-axes.
We will assume that RSD do not affect the void dimensions across the POS.
Hence, we can consider that $s_\perp = \rrs$.
An expression for $s_\parallel$, on the other hand, can be obtained by means of Eq.~(\ref{eq:halo_zspace}) adapted for the case of void-centric comoving distances:
\begin{equation}
    \Tilde{r}_\parallel = r_\parallel + v_\parallel \frac{(1 + \zsim)}{H(\zsim)},
	\label{eq:halo_void_zspace}
\end{equation}
where $r_\parallel$, $\Tilde{r}_\parallel$ and $v_\parallel$ are the void-centric analogues of $x_3$, $\Tilde{x}_3$ and $v_3$ in that equation.
In this case, $\Tilde{r}_\parallel$ and $r_\parallel$ must be replaced by $s_\parallel$ and $\rrs$ respectively.
The next ingredient is an expression for $v_\parallel$, which can be obtained from the void-centric velocity profile characterising the peculiar velocity field around voids.
This can be derived following linear theory via mass conservation up to linear order in density and assuming spherical symmetry \citep{vel_peebles, clues2, aprsd_hamaus1, aprsd_hamaus2, aprsd_correa}:
\begin{equation}
    v(r) = - \frac{1}{3} \frac{H(z)}{(1+z)} \beta(z) r \Delta(r),
	\label{eq:velocity}
\end{equation}
where $\Delta(r)$ is the integrated density contrast profile characterising the density field around voids, and $\beta(z)=f(z)/b$, the ratio between the logarithmic growth rate of density perturbations, $f(z)$, and the linear tracer-mass bias parameter, $b$.
In this profile, $v_\parallel = v(r_\parallel) = v(\rrs)$, for which $\Delta(r)$ must be evaluated at $r = \rrs$, which in turn is equal to the threshold of void identification: $\Delta(\rrs) = \Delta_{\rm id}$ (see Section~\ref{subsec:data_voids}, where we introduced the quantity $\Delta(r)$).
In this way, combining Eqs.~(\ref{eq:halo_void_zspace}) and (\ref{eq:velocity}) with the mentioned replacements, we get an expression for $s_\parallel$:
\begin{equation}
    s_\parallel = \rrs \left( 1 - \frac{1}{3} \beta(\zsim) \Delta_{\rm id} \right).
	\label{eq:q_elip}
\end{equation}
Note that here, we assumed the validity of hypotheses (2) and (3) concerning the isotropy of the density and velocity fields in r-space in order to explain a z-space phenomenon, even if this isotropy is no longer valid for z-space voids.

Our void finder identifies spherical regions instead of ellipsoidal zones.
Hence, as a first ansatz, we will assume that the z-space spherical voids enclose the same volume of the ellipsoids.
This is also depicted in the left panel of Fig.~\ref{fig:effects} with a red solid circle.
Calling $\rzs$ the radius of this equivalent sphere, equating both volumes, and using Eq.~(\ref{eq:q_elip}), it is straightforward to get an expression for $\rzs$:
\begin{equation}
    \rzs = \qrsd \rrs, \\
    \qrsd = \sqrt[3]{1 - \frac{1}{3} \beta(\zsim) \Delta_{\rm id}}.
	\label{eq:q1_rsd}
\end{equation}
Note that the theoretical factor $\qrsd$ is independent of the scale, it is only a constant of proportionality.
Moreover, it has a cosmological dependence, since it depends on $\beta$.
To get an explicit value for $\qrsd$, we need the corresponding values of $\Delta_{\rm id}$ and $\beta(\zsim)$.
As specified in Section~\ref{subsec:data_voids}, $\Delta_{\rm id} = -0.853$ for $\zsim = 0.99$.
The value of $\beta$ corresponding to the MXXL simulation was obtained following the procedure of \cite{aprsd_correa} by fitting Eq.~(\ref{eq:velocity}) to a measured r-space velocity profile with a Levenberg-Marquardt algorithm, getting in this way a value of ${\beta = 0.65}$.
With these two quantities, we have $\qrsd = 1.058$.
Note that $\qrsd > 1$, hence $\rzs > \rrs$, which is in agreement with our assumption that voids expand when they are mapped from r-space into z-space.

In reality, the actual value of $\rzs$ is expected to lay between $\rrs$ and $s_\parallel$ (Eq.~\ref{eq:q_elip}): $\rrs \leq \rzs \leq s_\parallel$.
In order to test for possible deviations in the predictions of Eq.~(\ref{eq:q1_rsd}), we also considered a more general approach by introducing a variable $\delta R_{\rm v}$ that quantifies the variation in radius:
\begin{equation}
    \delta R_{\rm v} := \frac{\rzs - \rrs}{s_\parallel - \rrs}.
    \label{eq:delta_R}
\end{equation}
Considering that $\qrsdc := \rzs/\rrs$ and combining Eqs.~(\ref{eq:delta_R}) and (\ref{eq:q_elip}), we obtain the following linear relation:
\begin{equation}
    \qrsdc = 1 - \frac{1}{3} \delta R_{\rm v} \beta(\zsim) \Delta_{\rm id}.
    \label{eq:q_rsd}
\end{equation}
Note that $\qrsdc \geq 1$ always, since $\Delta_{\rm id} < 0$.
Therefore, $\delta R_{\rm v}$ is expected to lay in the range $[0, 1]$.
In the limits, $\delta R_{\rm v} = 0$ ($\qrsdc = 1$) corresponds to the unlikely case of no expansion at all $\rzs = \rrs$, whereas $\delta R_{\rm v} = 1$ ($\qrsdc = 1.185$) corresponds to the case where the z-space void is characterized by a sphere with radius $\rzs=s_\parallel$.
This last case is also unlikely because it would mean that RSD affect the dimensions of voids equally in all directions.
In particular, the theoretical prediction of Eq.~(\ref{eq:q1_rsd}) corresponds to the case $\delta R_{\rm v}^s = 0.315$.

In Section~\ref{sec:stat} we will see that the theoretical prediction $\qrsd$ fits very well the median of the overall $\rzs/\rrs$ ratio distribution (see Fig.~\ref{fig:cor_radius}).
However, we found that larger voids respond better to the value $\delta R_{\rm v}^l = 0.5$, i.e., to a z-space radius that is the mean between $\rrs$ and $s_\parallel$.
We will discuss these aspects in more details in Sections~\ref{sec:stat} and \ref{sec:vsf}.
In this way, we get a prediction slightly different from Eq.~(\ref{eq:q1_rsd}):
\begin{equation}
    \rzs = \qrsdb \rrs, \\
    \qrsdb = 1 - \frac{1}{6} \beta(\zsim) \Delta_{\rm id},
	\label{eq:q2_rsd}
\end{equation}
with an explicit value of $\qrsdb = 1.092$.

The discrepancies between Eqs.~(\ref{eq:q1_rsd}) and (\ref{eq:q2_rsd}) can be attributed to the way in which the void finder performs the average spherical integration of the density field of an ellipsoidal underdense region.
Hence, the optimal value of $\qrsdc$ depends on the shape and slope of the real-space density profiles in the inner parts of voids.
We leave for a future investigation a deeper analysis about the derivation of $\qrsdc$ considering these aspects.


\subsection{Alcock-Paczy\'nski change of volume}
\label{subsec:effects_ap}

Up to here, a true distance scale was implicitly assumed.
Note however that the only information available from observational catalogues are angular positions and redshifts of astrophysical objects like galaxies.
These observables must be transformed into a Mpc-scale, which involves the use of a fiducial cosmology.
A deviation between the true and fiducial cosmologies will lead to additional distortions in the spatial distribution of galaxies.
This is a manifestation of the AP effect, and will also affect the volume of voids.
To understand this effect, let us consider the distribution of galaxies in r-space (free of RSD) for the following analysis.

The size of a spherical void can be quantified by a POS angular radius $\Delta \theta$, and a LOS redshift radius $\Delta z$.
These observables are related to physical dimensions ($R_{\rm v \perp}, R_{\rm v \parallel}$)
by the following transformation equations:
\begin{equation}
    R_{\rm v \perp} = D_{\rm M}(\zsim) \Delta \theta, \\
    R_{\rm v \parallel} = \frac{d D_{\rm M}}{dz}(\zsim) \Delta z = \frac{c}{H(\zsim)} \Delta z,
	\label{eq:void_sizes}
\end{equation}
where $D_{\rm M}$ is the comoving angular diameter distance.
Hence, the pair of Eqs.~(\ref{eq:void_sizes}) depend on cosmology through the Hubble parameter (Eq.~\ref{eq:hubble}).
Note that if one knew the true cosmology, then it would not be necessary to distinguish between the POS and LOS dimensions.
Both would be equal to the r-space void radius: $\rrs = R_{\rm v \parallel} = R_{\rm v \perp}$.
However, assuming a fiducial cosmology leads to discrepancies between both quantities, and a spherical void will appear as an ellipsoid in the underlying coordinate system.
Nevertheless, unlike the RSD-ellipsoids from the expansion effect, the AP-ellipsoids are distorted in both the POS and LOS directions.
Even more, the net result is not necessary an expansion, it can also be a contraction, it all depends on the chosen cosmology.
This additional \textit{AP change of volume} is depicted schematically in the middle panel of Fig.~\ref{fig:effects}.

We can describe these AP distortions following a similar approach to that used for the expansion effect.
Considering that the AP-ellipsoid has semi-axes $(s_\perp^{\rm fid}, s_\parallel^{\rm fid})$, given by Eqs.~(\ref{eq:void_sizes}) with fiducial values $H_{\rm fid}$ and $D_{\rm M}^{\rm fid}$, then a direct comparison with ($R^{\rm true}_{\rm v \perp}, R^{\rm true}_{\rm v \parallel}$) leads to the following relations:
\begin{equation}
    s_\perp^{\rm fid} = \qap^\perp R^{\rm true}_{\rm v \perp}, \\
    s_\parallel^{\rm fid} = \qap^\parallel R^{\rm true}_{\rm v \parallel},
    \label{eq:q_ap_pos_los2}
\end{equation}
where
\begin{equation}
    \qap^\perp = \frac{D_{\rm M}^{\rm fid}(\zsim)}{D_{\rm M}^{\rm true}(\zsim)}, \\
    \qap^\parallel = \frac{h_{\rm true}\varepsilon_{\rm true}(\zsim)}{h_{\rm fid}\varepsilon_{\rm fid}(\zsim)}.
	\label{eq:q_ap_pos_los}
\end{equation}
Here, $\varepsilon(z)$ is the square root term of Eq.~(\ref{eq:hubble}), and we adopted the index "true" to refer to quantities based on the true underlying cosmology.
Finally, considering the equivalent sphere with the same volume of the ellipsoid, and calling $R_{\rm v}^{\rm fid}$ this new radius, we get an expression similar to Eq.~(\ref{eq:q1_rsd}):
\begin{equation}
    R_{\rm v}^{\rm fid} = \qap \rrs, \\
    \qap = \sqrt[3]{(\qap^\perp)^2 \qap^\parallel}.
	\label{eq:q_ap}
\end{equation}

Like the RSD factor, $\qap$ is also a constant of proportionality and cosmology dependent.
However, there is an interesting difference between them.
The AP factor depends only on the background cosmological parameters, whereas the RSD factor depends only on $\beta$.
Therefore, $\qap$ encodes information about the expansion history and geometry of the Universe, whereas $\qrsdc$, about the growth rate of cosmic structures.

In the development of Section~\ref{sec:vsf} we will need explicit values of the $\qap$ factors for the FC void catalogues defined in Section~\ref{subsec:data_voids}.
For the FC-l catalogue, with $\Omega_m^l = 0.20$, we get a value of $\qap^l = 1.046$.
For the FC-u catalogue, with $\Omega_m^u = 0.30$, a value of $\qap^u = 0.960$.
Note that $\qap^l > 1$ for the former, hence according to Eq.~(\ref{eq:q_ap}) it is expected an expansion of the fiducial voids.
Conversely, $\qap^u < 1$ for the latter, hence it is expected a contraction.


\subsection{Combined AP and RSD contributions}
\label{subsec:effects_ap_rsd}

The volume of a void will be affected by the combined contributions of the AP and RSD effects, which are indistinguishable in observations.
A priori, it is not trivial to ensure that both effects can be treated independently as we did.
However, in Section~\ref{sec:vsf} we will provide evidence of this.
From a theoretical point of view, the fact that the $\qrsdc$ and $\qap$ factors encode different cosmological information is a good sign of this assumption.

Assuming this independence, we can relate the z-space and r-space void radii making a two-step correction: first, we apply Eq.~(\ref{eq:q_ap}) to correct for the AP effect, and then, we apply Eq.~(\ref{eq:q1_rsd}) (or Eq.~\ref{eq:q2_rsd}) to correct for the RSD expansion effect:
\begin{equation}
    \rzs = \qap ~ \qrsdc ~ \rrs.
    \label{eq:q_ap_rsd}
\end{equation}


\subsection{Off-centring effect}
\label{subsec:effects_offcentring}

A simple visual inspection of Fig.~\ref{fig:mxxl_slice} shows that z-space void centres are shifted with respect to their r-space counterparts.
This \textit{off-centring} is a direct consequence of the failure of hypothesis (4) concerning the invariability of centre positions when voids are mapped from r-space into z-space.
\cite{reconstruction_nadathur} remarks that this hypothesis is equivalent to assuming that void positions do not suffer RSD themselves.
On the other hand, \citet{lambas_sparkling_2016}, \citet{ceccarelli_sparkling_2016} and \citet{lares_sparkling_2017} demonstrated that voids move as whole entities with a net velocity $\boldsymbol{V}_{\rm v}$.
Inspired by these results, then the off-centring effect can be simply understood as a new kind of RSD induced by void dynamics, and therefore it is expected that void centres appear shifted along the LOS when they are identified in z-space, in the same way as tracers do.
The right panel of Fig.~\ref{fig:effects} depicts schematically this effect.
We will provide a solid evidence of this effect in Section~\ref{subsec:stat_desp_vel}.

We can make an analytical prediction of this effect considering it as a dynamical phenomenon.
The void finder used in this work provides the positions $\posv = (\possvx, \possvy, \possvz)~[\hmpc]$ and peculiar velocities $\velv = (\velvx, \velvy, \velvz)~[\kms]$ of void centres (see Section~\ref{sec:stat} for more details about how the velocities were calculated).
Therefore, in order to account for the LOS shifting of centres, it is only necessary to write an expression equivalent to Eq.~(\ref{eq:halo_zspace}) for voids:
\begin{equation}
    \Tilde{X}_{\rm v 3} = \possvz + \velvz \frac{(1 + \zsim)}{H(\zsim)},
	\label{eq:void_zspace}
\end{equation}
where $\Tilde{X}_{\rm v3}$ denotes the shifted $\possvz$-coordinate.
As was the case of Section~\ref{subsec:effects_ap_rsd}, it is not trivial to know if the expansion and off-centring effects are independent from each other.
In Section~\ref{subsec:stat_cross} we will provide evidence of this.
In this way,  Eq.~(\ref{eq:halo_void_zspace}) is still valid, provided that $r_\parallel = |\possvz - x_3|$ and $v_\parallel = |\velvz - v_3|$.

Note that the expansion effect is a by-product of RSD induced by \textit{tracer dynamics} at scales around the void radius. 
At these scales, the velocity field of tracers responds to a divergence originated in the local density minimum located at the void.
On the other hand, the off-centering effect is a result of RSD induced at larger scales.
Its source is the bulk motion of galaxy tracers in the whole region containing the void following the large scale dynamics of the gravitational field \citep{lares_sparkling_2017}.
This last aspect resembles a \textit{void dynamics}, that presents itself as a different kind of RSD.
Therefore, it is expected that both effects leave a footprint on cosmological statistics such as the void size function and the void-galaxy correlation function.
Hereafter, we will refer to both as the t-RSD expansion effect and the v-RSD off-centring effect, hinting at their different nature: t for tracer dynamics, and v for void dynamics.


\section{Statistical analysis}
\label{sec:stat}

The statistical analysis of this section aims to provide evidence about the t-RSD expansion effect and the v-RSD off-centring effect introduced in last section.
In next section, we will complete the analysis incorporating the additional AP change of volume.
For this reason, we will continue using the TC catalogues, specifically the bijective samples (TC-rs-b and TC-zs-b in Table~\ref{tab:catalogues}).

This analysis is based on correlations between three statistics that characterise the volume alteration and movement of a void:
(i) z-space to r-space radius ratio $\q$, (ii) z-space displacement of the centre $\disp = (\posvx,\posvy,\posvz)$, and (iii) r-space net velocity $\velv = (\velvx,\velvy,\velvz)$.
Specifically, $\disp$ was calculated as the displacement of a void centre in going from r-space into z-space normalised to the r-space radius:
\begin{equation}
    \disp = \frac{\Tilde{\boldsymbol{X}}_{\rm v} - \posv}{\rrs},
    \label{eq:disp_centres}
\end{equation}
whereas $\velv$ was computed summing all the individual velocities of haloes inside a spherical shell with dimensions $0.8 \leq r/\rrs \leq 1.2$.
This velocity is an unbiased and fair estimation of the bulk flow velocity of the void, as was demonstrated in \citet{lambas_sparkling_2016} (see their Fig.~1).


\subsection{Correlations between r-space and z-space void radii}
\label{subsec:stat_radius}

\begin{figure*}
    \includegraphics[width=\columnwidth]{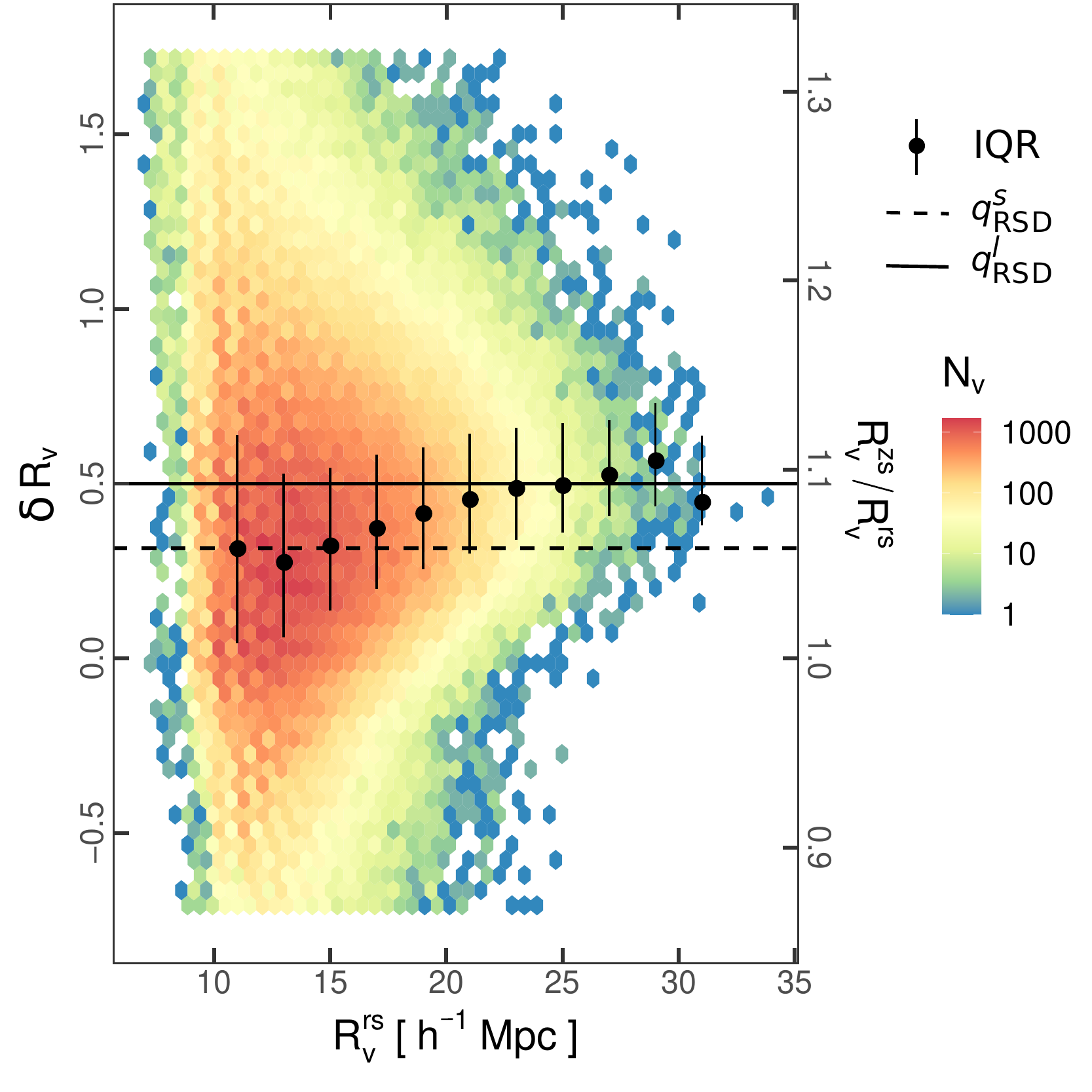}
    \includegraphics[width=\columnwidth]{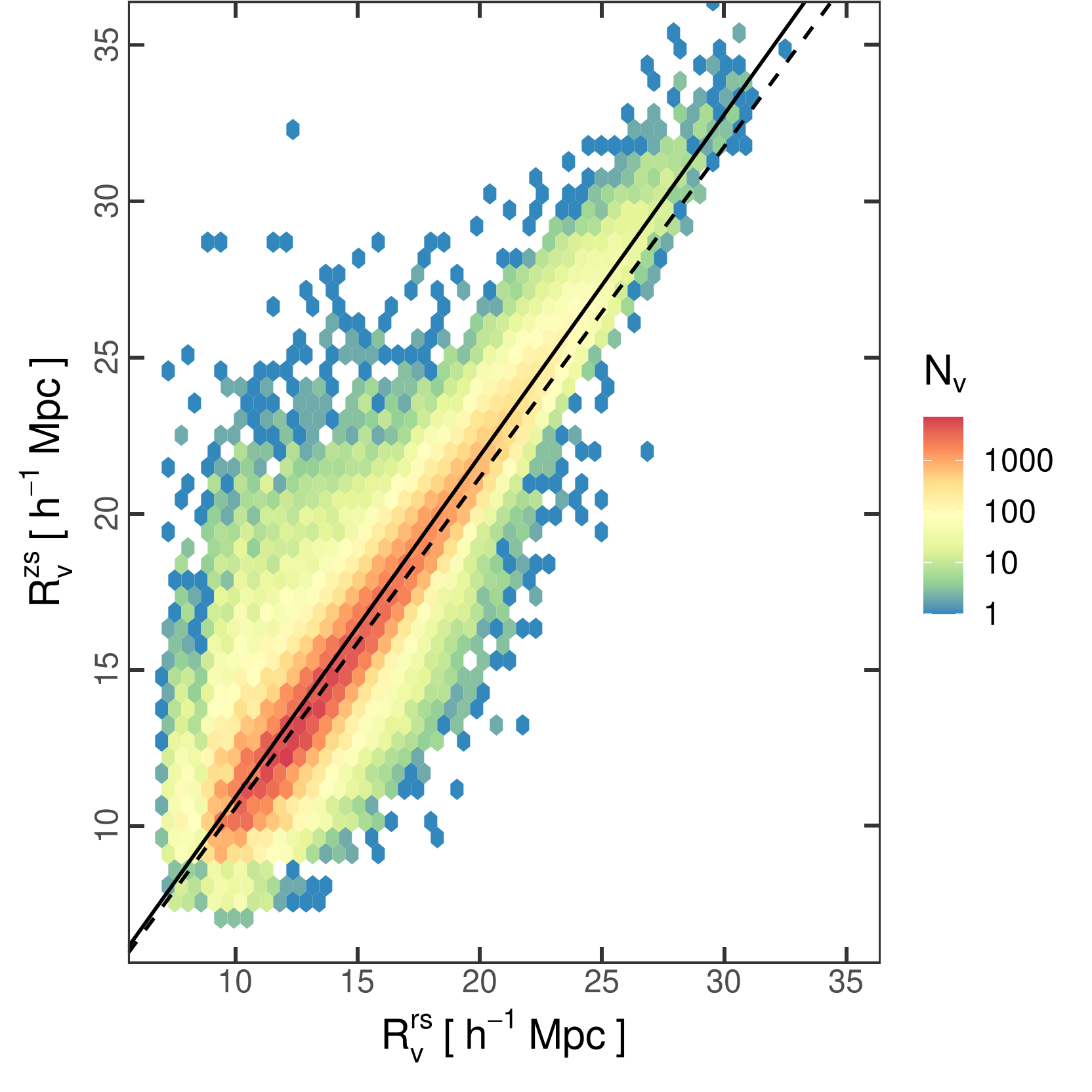}
    \caption{
    \textit{Left panel.}
    2D distribution $(\rrs,\dRc)$ shown as a heat map.
    From blue to red, the colours span from low to high number of void counts in a logarithmic scale.
    The right axis shows the equivalent scale based on the ratio $\q$, related to $\dRc$ via Eq.~(\ref{eq:q_rsd}).
    The black dots with error bars show the evolution of the median and interquartile range (IQR) of this distribution with void radius.
    The horizontal dashed and solid lines indicate the predictions of Eqs.~(\ref{eq:q1_rsd}) ($\dR = 0.315$, $\qrsd = 1.058$) and (\ref{eq:q2_rsd}) ($\dRb = 0.5$, $\qrsdb = 1.092$) respectively.
    \textit{Right panel.}    
    2D distribution $(\rrs,\rzs)$.
    There is a linear trend, whose slope is correctly described by the RSD factors $\qrsd$ (dashed line) and $\qrsdb$ (solid line).
    In both panels, it is clear that $\qrsd$ is a better predictor of the median for smaller voids, whereas $\qrsdb$ is more suitable for the larger ones.
    This analysis constitutes a statistical demonstration of the t-RSD expansion effect.
    }
    \label{fig:cor_radius}
\end{figure*}

The left panel of Fig.~\ref{fig:cor_radius} shows the 2D distribution $(\rrs, \dRc)$ as a heat map.
From blue to red, the colours span from low to high void counts $N_{\rm v}$.
These counts are presented in a logarithmic scale in order to highlight the patterns of the distribution at different scales.
The right axis shows the equivalent scale based on the ratio $\q$, related to $\dRc$ via Eq.~(\ref{eq:q_rsd}).
In order to study the evolution of this distribution with void radius, we computed the median and interquartile range (IQR) taking bins of width $2~\hmpc$ in the range $10 \leq \rrs/\hmpc \leq 32$.
This is shown by black dots with error bars.
The horizontal dashed and solid lines indicate the predictions of Eqs.~(\ref{eq:q1_rsd}) ($\dR = 0.315$, $\qrsd = 1.058$) and (\ref{eq:q2_rsd}) ($\dRb = 0.5$, $\qrsdb = 1.092$) respectively.
Note that $\qrsd$ is a better predictor of the median for smaller voids, whereas $\qrsdb$ is more suitable for larger voids, the ones more interesting for cosmological studies.

The right panel of Fig.~\ref{fig:cor_radius} shows the 2D distribution $(\rrs, \rzs)$.
There is a clear linear trend between both radii, whose slope can be correctly described by the RSD factors $\qrsd$ (dashed line) and $\qrsdb$ (solid line).
As before, $\qrsdb$ is more suitable for larger voids.

From this analysis, we arrive at the second important conclusion of this work: voids expand when they are mapped from r-space into z-space, and this expansion can be statistically quantified as an increment in void radius by a factor $\qrsdc$
($\qrsd$ for smaller voids or $\qrsdb$ for larger voids).
These results give support to the t-RSD expansion effect postulated in Section~\ref{subsec:effects_expansion}.

\subsection{Correlations between displacement of centres and net velocity }
\label{subsec:stat_desp_vel}

\begin{figure}
    \includegraphics[width=\columnwidth]{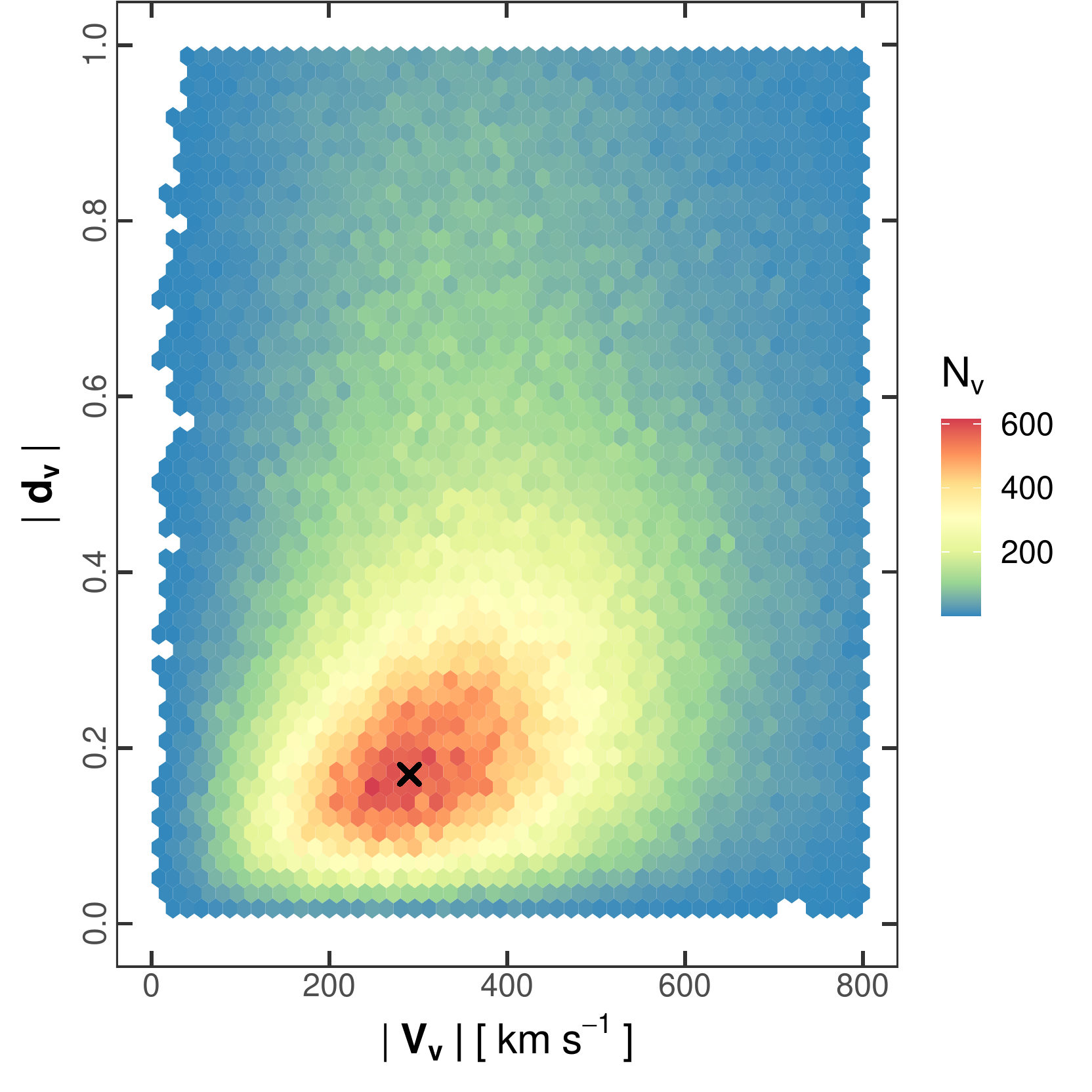}
    \caption{
    2D distribution $(|\velv|, |\disp|)$.
    The cross indicates the 2D mode, which shows that voids tend to move with a speed of $290~\kms$, and their centres tend to shift an amount of $0.17~\rrs$.
    }
    \label{fig:cor_disp_vel}
\end{figure}

Fig.~\ref{fig:cor_disp_vel} shows the 2D distribution $(|\velv|, |\disp|)$, where the moduli of these vectors were taken.
This distribution contains information about the dynamics of voids as whole entities.
The cross in the figure indicates the mode of the 2D distribution, which shows that voids tend to move with a speed of $290~\kms$, and their centres tend to displace an amount of $0.17~\rrs$.
It is clear then, that voids can not be considered at rest.

\begin{figure}
	\includegraphics[width=\columnwidth]{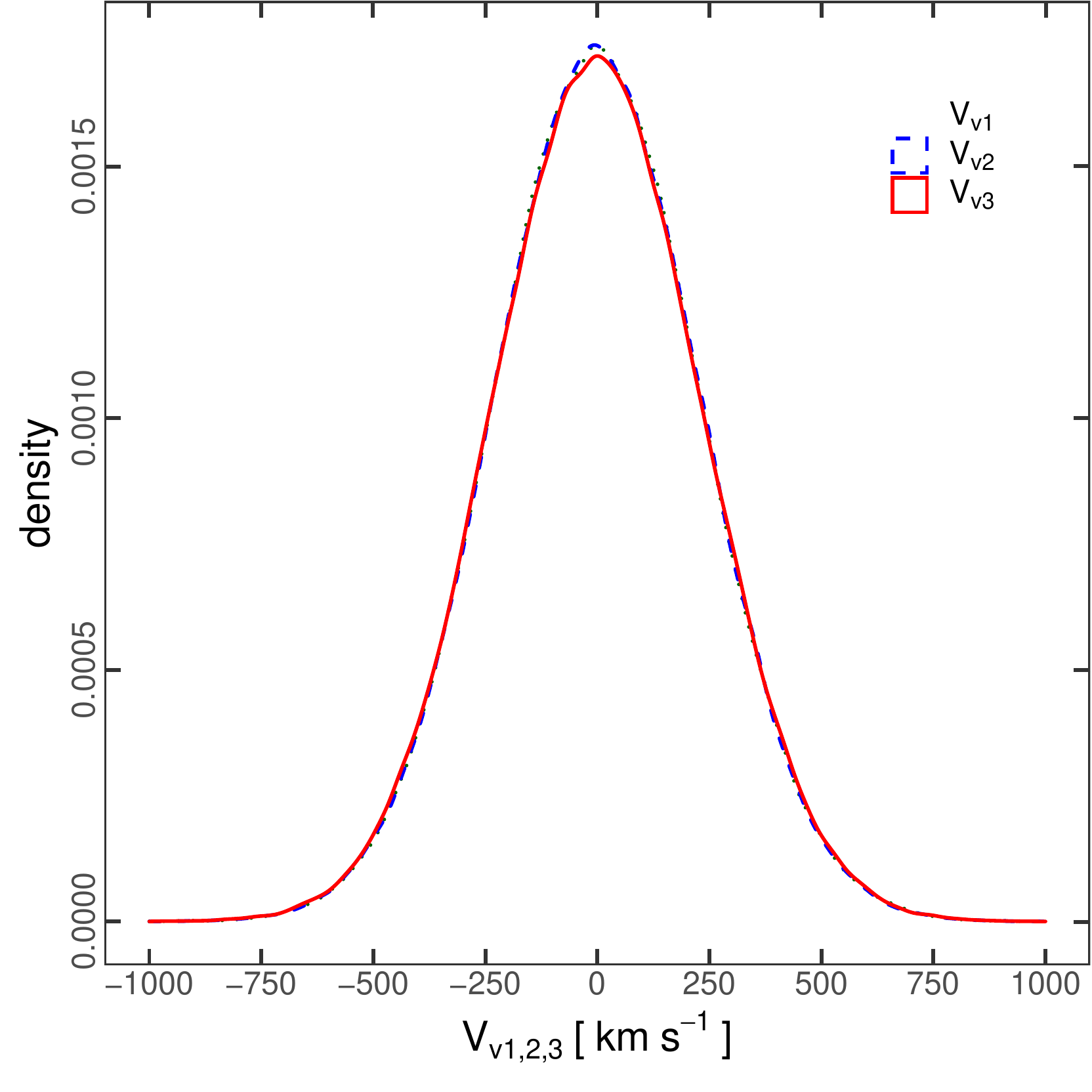}
    \caption{
    Distribution of the components of $\velv$ along the three directions of the simulation box.
    They show Gaussian shapes, centred at $0~\kms$, with a dispersion of $231~\kms$.
    This is a manifestation of the isotropic movement of voids.
    }
    \label{fig:hist_vel}
\end{figure}

Concerning the velocities, Fig.~\ref{fig:hist_vel} shows the distribution of the components of $\velv$ along the three directions of the simulation box.
They show Gaussian shapes, centred at $0~\kms$, with a dispersion of $231~\kms$.
This was expected, since there is not any privileged direction of motion for voids in simulations.

\begin{figure*}
	\includegraphics[width=\columnwidth]{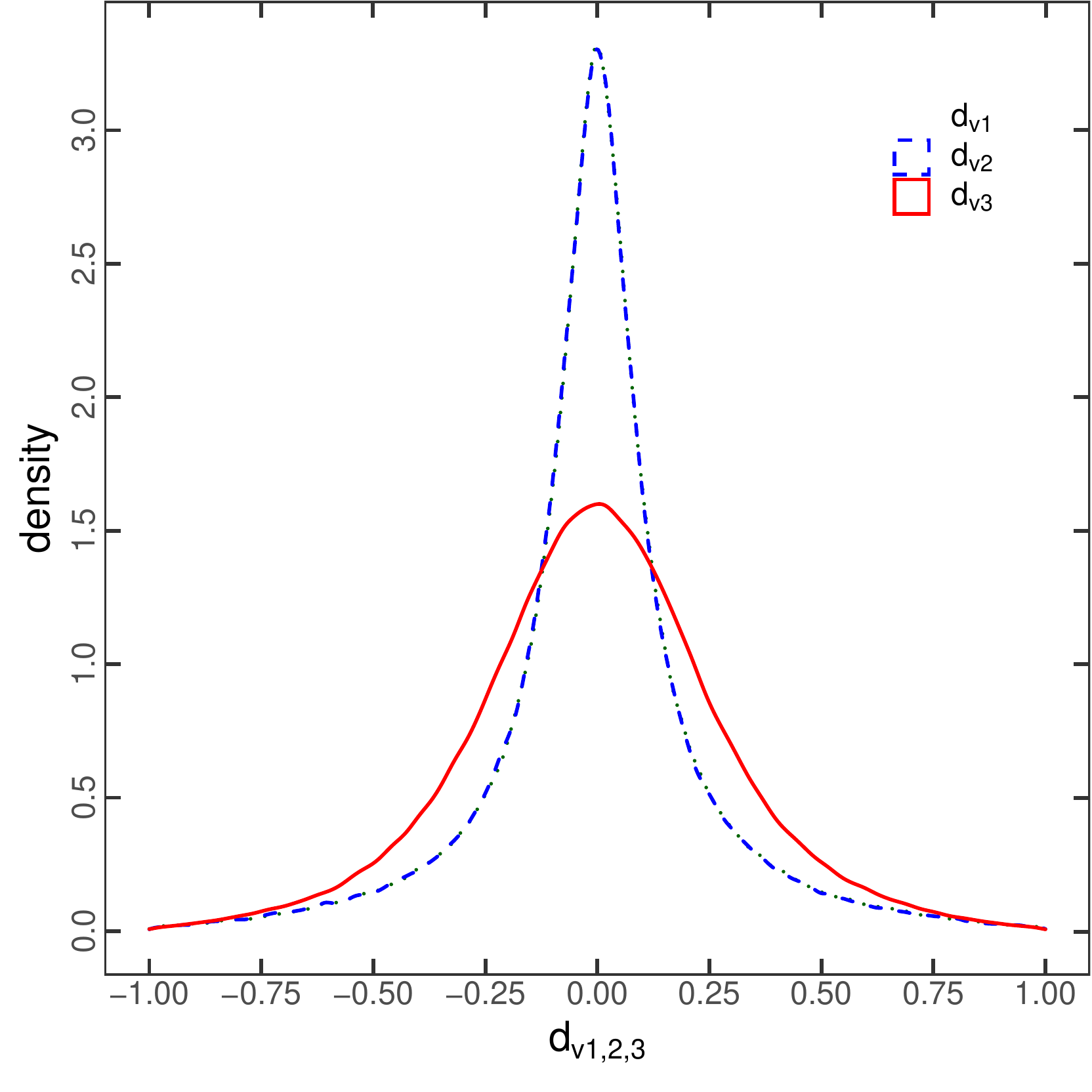}
	\includegraphics[width=\columnwidth]{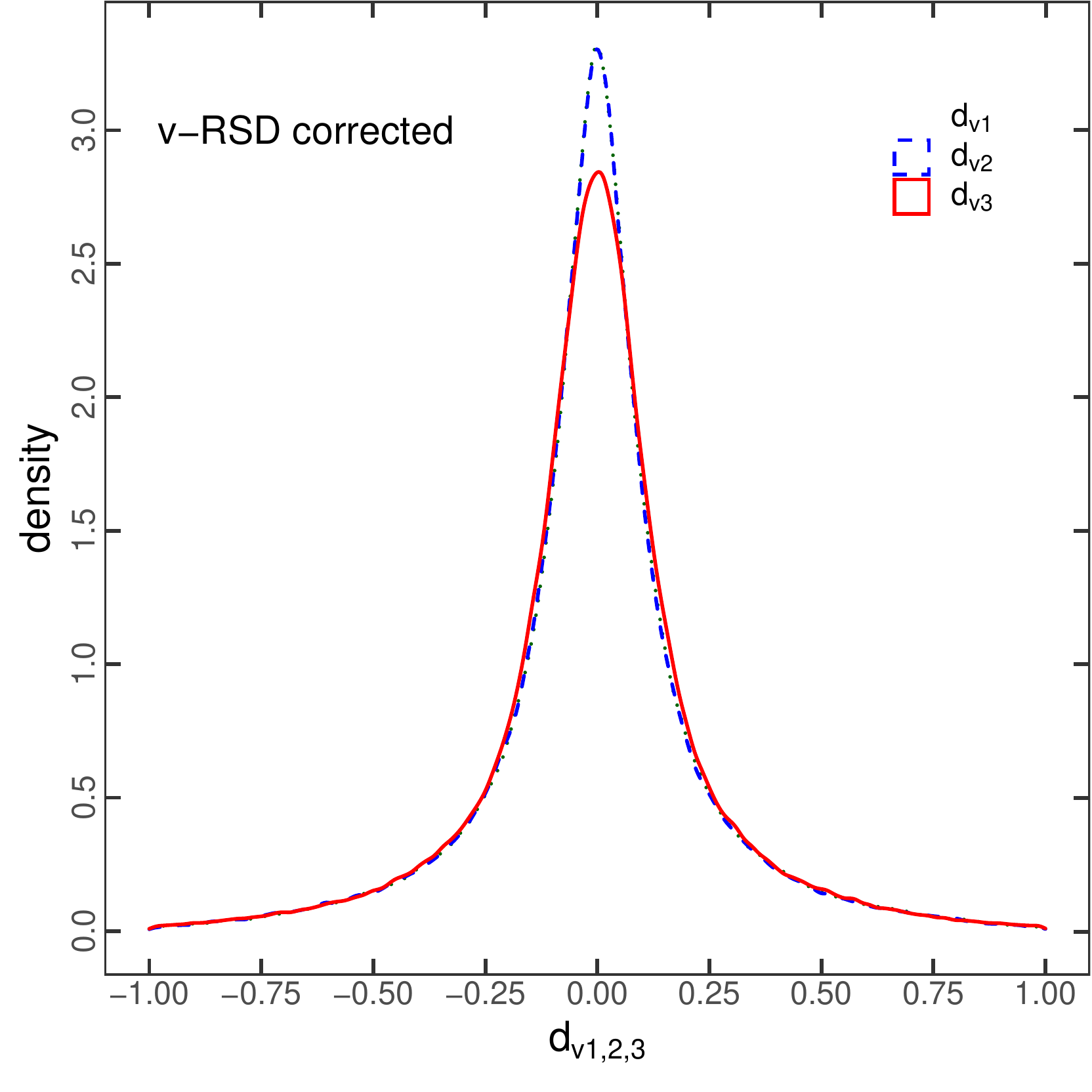}
    \caption{\textit{Left panel.}
    Distribution of the components of $\disp$ along the three directions of the simulation box.
    They show Gaussian shapes centred at $0$.
    The POS distributions (green dotted and blue dashed lines) are almost identical with a dispersion of $0.25$, whereas the LOS distribution (red solid line) is very different, with a dispersion of $0.3$.
    \textit{Right panel.}
    The three distributions are almost identical after correcting the LOS displacements with Eq.~(\ref{eq:void_zspace}).
    This is a statistical demonstration of the v-RSD off-centring effect.
    }
    \label{fig:hist_disp}
\end{figure*}

Concerning the displacements, the left panel of Fig.~\ref{fig:hist_disp} shows the distribution of the components of $\disp$ along the three directions of the simulation box.
They all show Gaussian shapes centred at $0$.
However, unlike velocities, displacements are different depending on the direction.
On the one hand, the POS distributions (green dotted and blue dashed lines) are almost identical, as expected, with a dispersion of $0.25$.
On the other hand, the LOS distribution (red solid line) has a dispersion of $0.3$.
Nevertheless, after correcting the LOS displacements with Eq.~(\ref{eq:void_zspace}), the POS distribution is recovered, as shown in the right panel.
These distributions reflect a residual and isotropic displacement that can be attributed to Poisson noise when the void finder tries to localise the optimum centre (step iv in Section~\ref{subsec:data_voids}).

\begin{figure*}
    \includegraphics[width=\columnwidth]{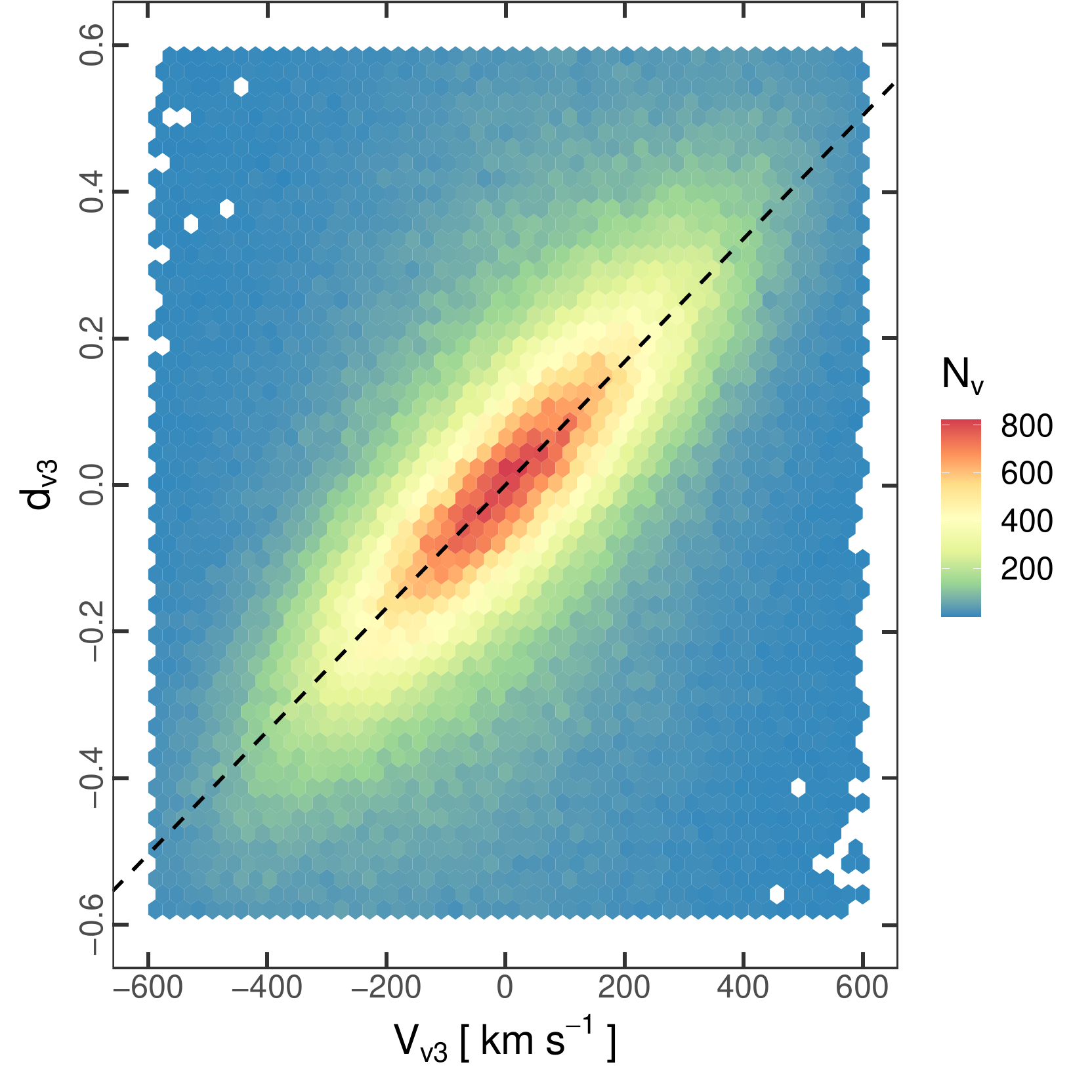}
    \includegraphics[width=\columnwidth]{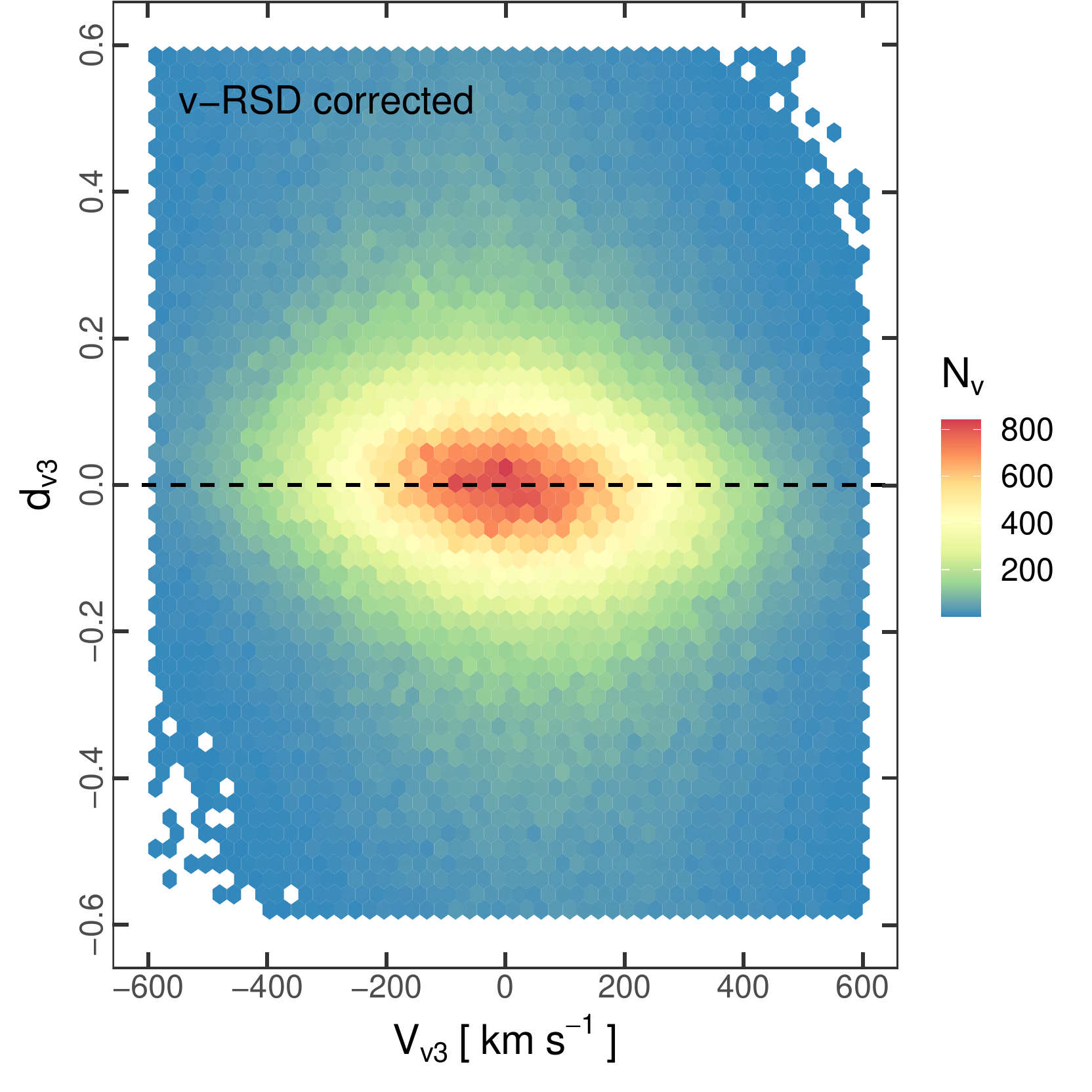}
    \caption{
    \textit{Left panel}.
    2D distribution $(\velvz, \posvz)$.
    There is a linear trend that is correctly described by the linear relation given by Eq.~(\ref{eq:void_zspace}) (dashed line).
    \textit{Right panel}.
    The correlation disappears after correcting the LOS displacements with this equation.
    This is a statistical demonstration of the v-RSD off-centring effect.
    }
    \label{fig:cor_disp3_vel3}
\end{figure*}

The phenomenon described in the last paragraph is more evident in the left panel of Fig.~\ref{fig:cor_disp3_vel3}, where the 2D LOS distribution $(\velvz, \posvz)$ is shown.
There is a linear trend between both quantities, which is correctly described by Eq.~(\ref{eq:void_zspace}) represented by the dashed line.
Specifically, the slope of this line is given by the term $(1+\zsim)/H(\zsim)$.
The right panel of Fig.~\ref{fig:cor_disp3_vel3} shows that after correcting the LOS displacements with this equation, the correlation disappears, leading to a 2D distribution that is almost identical to the corresponding ones of the POS components, $(\velvx, \posvx)$ and $(\velvy, \posvy)$ (not shown here).

From this analysis, we arrive at the third important conclusion of this work: void centres shift preferentially along the LOS when they are mapped from r-space into z-space, and this displacement can be statistically quantified by means of Eq.~(\ref{eq:void_zspace}).
These results give support to the v-RSD off-centring effect postulated in Section~\ref{subsec:effects_offcentring}.


\subsection{Cross correlations}
\label{subsec:stat_cross}

\begin{figure*}
    \includegraphics[width=\columnwidth]{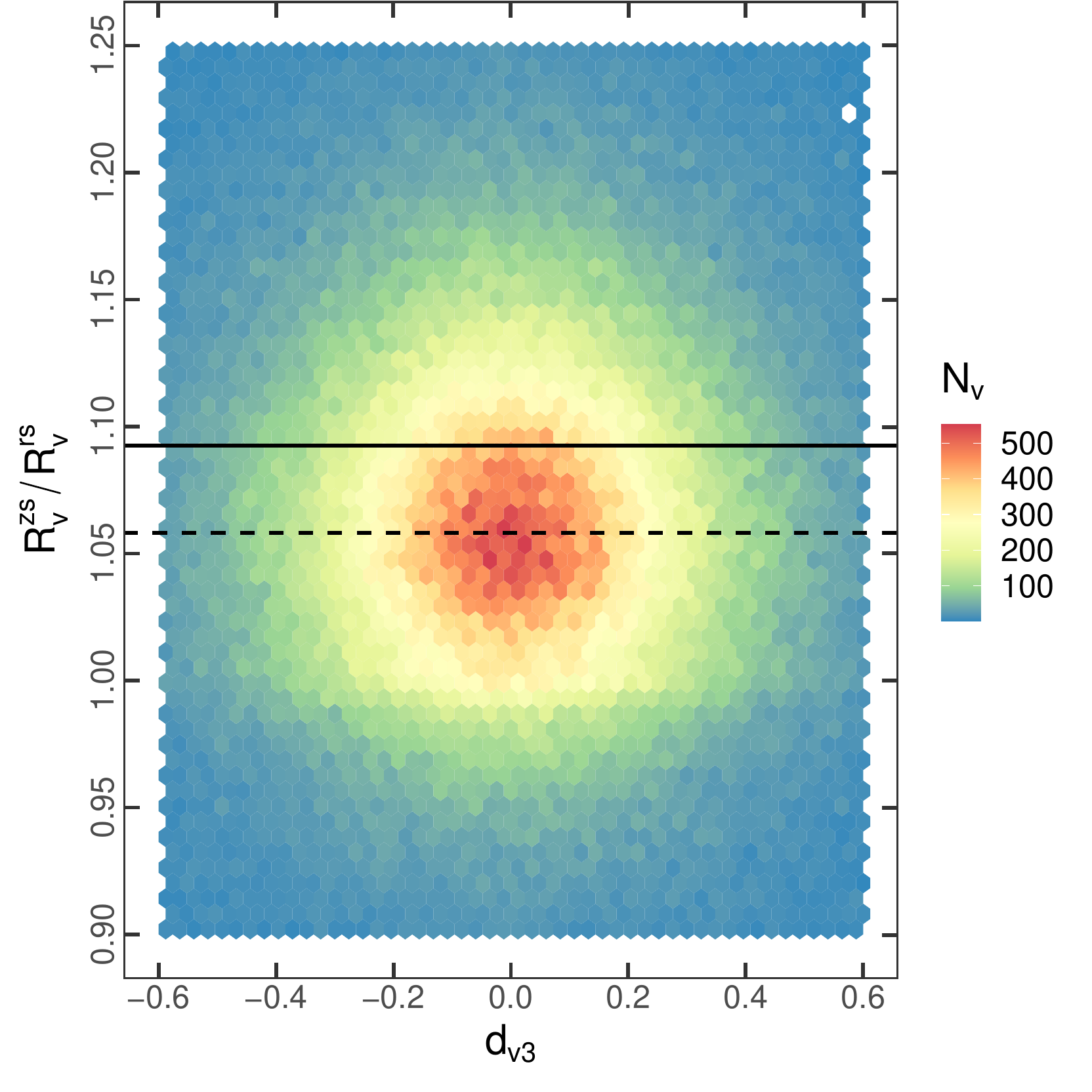}
    \includegraphics[width=\columnwidth]{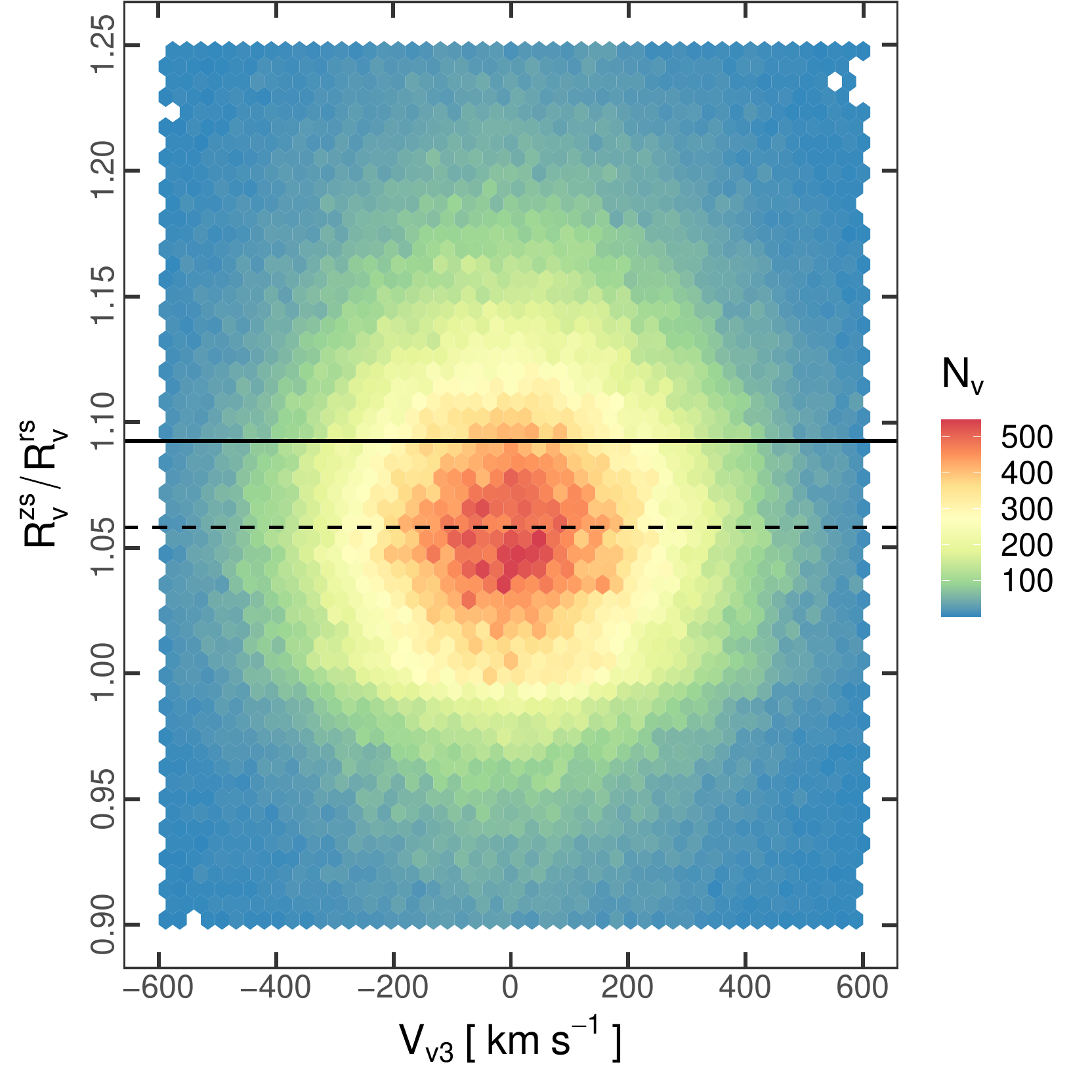}
    \caption{
    2D distributions $(\posvz, \q)$ (left panel) and $(\velvz, \q)$ (right panel).
    The horizontal lines are the theoretical predictions $\qrsd$ (Eq,~\ref{eq:q1_rsd}, dashed) and $\qrsdb$ (Eq.~\ref{eq:q2_rsd}, solid).
    No correlations can be seen in both panels, suggesting that the t-RSD expansion effect and the v-RSD off-centring effect are statistically independent.
    }
    \label{fig:cor_cross}
\end{figure*}

Since the ratio $\q$ (or equivalently $\dRc$) characterises the change of volume in voids, the statistical analysis of Section~\ref{subsec:stat_radius} gave support to the t-RSD expansion effect postulated in Section~\ref{subsec:effects_expansion}.
On the other hand, as the displacement $\disp$ and velocity $\velv$ characterise the movement of voids, the statistical analysis of Section~\ref{subsec:stat_desp_vel} gave support to the v-RSD expansion effect postulated in Section~\ref{subsec:effects_offcentring}.
It only remains to test if both effects are statistically independent by looking for cross correlations between these quantities.
Fig.~\ref{fig:cor_cross} shows the 2D LOS distributions $(\posvz, \q)$ (left panel) and $(\velvz, \q)$ (right panel).
The horizontal lines are the theoretical predictions $\qrsd$ (dashed) and $\qrsdb$ (solid).
No correlations can be seen, giving support to the postulated independence.
It is worth mentioning that the analogue POS distributions (not shown here): $(\posvx, \q)$, $(\posvy, \q)$, $(\velvx, \q)$ and $(\velvy, \q)$ show a similar behaviour.
These results allow us to interpret the large scale dynamics of the whole region containing the void (v-RSD) as decoupled from the dynamics of the galaxies at scales of the void radius (t-RSD).
This also suggests that potential distortion patterns in observations due to these two effects can be treated separately.
This is the fourth important conclusion of this work.

\section{Impact on the void size function}
\label{sec:vsf}

This section has a double intention.
On the one hand, we will finish the analysis of the last section incorporating the additional AP change of volume not treated yet.
On the other hand, we will study the impact of all the z-space effects on the void size function.
For this reason, we turn now to the FC void catalogues, fully affected by z-space systematics (see Table~\ref{tab:catalogues}).


\subsection{Generalities of the VSF modelling}
\label{subsec:vsf_vsf}

We start with some generalities concerning the VSF modelling.
The real-space VSF can be modelled using the excursion set formalism combined with the spherical expansion of matter underdensities derived from perturbation theory \citep{SvdW}.
This model is analogous to that used to describe the abundance of dark matter haloes.
Contrary to the case of haloes, which collapse, voids expand (void-in-void mode).
Hence, this formalism uses an underdense barrier to take into account the abundance of voids.
This value is taken from the moment of shell-crossing in the expansion process.
However, this is not the complete picture.
There are some voids embedded in an overdense shell shrinking due to gravitational collapse.
This is the void-in-cloud mode.
Therefore, in order to take into account this mode too, the overdense barrier corresponding to collapse is also added in the excursion set theory.

There are two main approaches for this modelling: the \citet[SvdW]{SvdW} model, and the \citet[Vdn]{jennings_Vdn} model.
The key assumption of the former is isolated expansion, namely, it assumes that the comoving number density is conserved during expansion.
However, this leads to a cumulative volume fraction of voids that exceeds unity.
In order to fix this problem, the latter assumes that, instead of the number density, the comoving volume fraction is conserved during expansion.

It is important to highlight that both models are only applicable to dark matter voids.
Halo or galaxy voids are substantially different in their statistical properties.
Nevertheless, many authors claim that both types of voids can still be related to each other with a linear bias approach
%
%
%
\citep{abundance_furlanetto,bias_pollina1,bias_pollina2,bias_chan2,abundance_bias_contarini,density_bias_fang,bias_pollina3,bias_schuster,bias_chan3}, 
and hence, the SvdW and Vdn models are still valid.
Such a model should fit the r-space abundances of Fig.~\ref{fig:TC_VSF} (blue lines).

In practice, z-space galaxy voids are used in observations.
Therefore, the z-space systematics proposed in this work are expected to have a strong impact on the VSF.
We will tackle this problematic using the theoretical machinery developed in Section~\ref{sec:effects}.
As we explained there, this framework depends strongly on cosmology, hence it must be combined with the excursion set formalism in order to obtain unbiased cosmological constraints from redshift surveys.
In this way, we lay the foundations for a complete treatment of the VSF modelling, leaving for a future investigation a full analysis combining both developments.


\subsection{Alcock-Paczy\'nski correction}
\label{subsec:vsf_AP}

\begin{figure*}
    \includegraphics[width=\columnwidth]{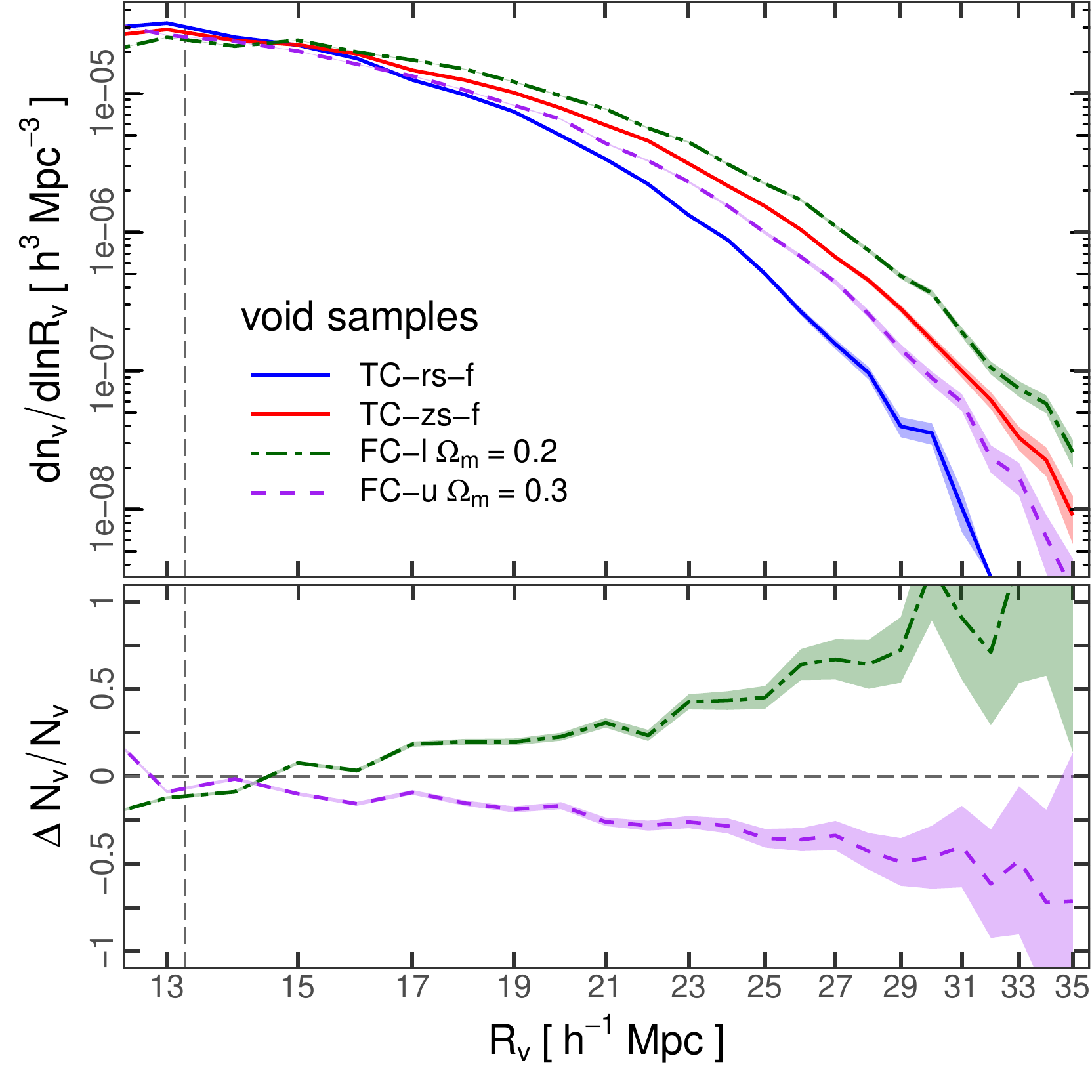}
    \includegraphics[width=\columnwidth]{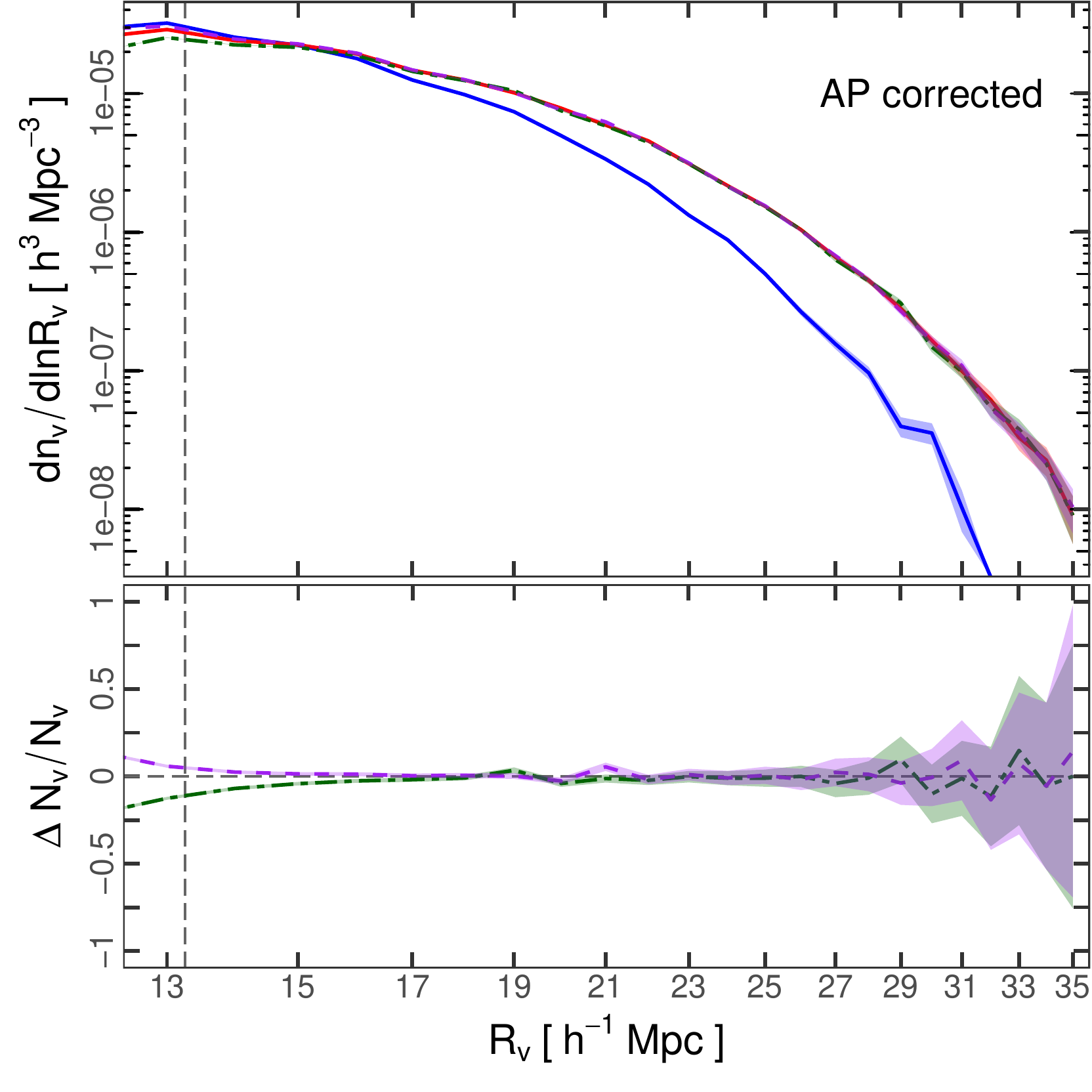}
    \caption{Alcock-Paczy\'nski correction of void abundances.
    \textit{Left panel.}
    VSFs of the FC void samples.
    The VSF of the FC-l sample, which assumes a fiducial value of $\Omega_m^l = 0.20$, is represented with a green dot-dashed line, whereas the VSF of the FC-u sample, which assumes a fiducial value of $\Omega_m^u = 0.30$, is represented with a purple dashed line.
    By way of comparison, the VSF of the TC full r-space and z-space samples (blue and red solid lines respectively) are also shown as references.
    The corresponding fractional differences of void counts between the FC samples and the TC z-space sample are shown in the lower panel.
    \textit{Right panel.}
    The same as the left panel, but after correcting the FC abundances for the AP change of volume.
    }
    \label{fig:FC_VSF_AP}
\end{figure*}

The left panel of Fig.~\ref{fig:FC_VSF_AP} shows the void abundances of the two FC void samples, which are fully affected by z-space systematics, and hence, mimic two possible observational measurements.
The VSF of the FC-l sample, which assumes a lower fiducial value with respect to that of the simulation ($\Omega_m^l = 0.20$) is represented with a green dot-dashed line, whereas the VSF of the FC-u sample, which assumes an upper fiducial value ($\Omega_m^u = 0.30$), with a purple dashed line.
The goal of this section is to verify if the theoretical framework developed in Section~\ref{sec:effects} allows to correct these abundances curves in order to recover the true underlying r-space one, unaffected by any of the z-space systematics (blue solid line in the plot, corresponding to the TC-rs-f sample).
To do this, it is sufficient to correct each void radius just applying Eq.~(\ref{eq:q_ap_rsd}), using the values of the AP and RSD factors that we have derived: $\qrsdb=1.092$, $\qap^l=1.046$ and $\qap^u=0.960$.

Instead of performing this correction directly, we will split it in a two-step procedure in order to discuss the different physical mechanisms involved.
In this subsection, we discuss the first step, correcting for the AP change of volume with the AP factors.
In the next subsection, we will discuss the second step, correcting for the t-RSD expansion effect with the RSD factor.
Therefore, the goal of this subsection is to recover the z-space VSF which is affected by RSD but unaffected by the AP effect (red solid line in the plot, corresponding to the TC-zs-f sample).
This correction is shown in the right panel of Fig.~\ref{fig:FC_VSF_AP}.

The first aspect clearly seen in the left panel of Fig.~\ref{fig:FC_VSF_AP} when comparing the abundances of the FC-l and FC-u samples with respect to the z-space VSF of reference, is that a higher VSF is obtained when a lower value of $\Omega_m$ is assumed, whereas the opposite behaviour occurs when a higher value of $\Omega_m$ is assumed.
In the context of the bijective mapping, this means that the FC-l voids are systematically bigger, whereas the FC-u voids are smaller.
This is in agreement with our discussion in Section~\ref{subsec:effects_ap}, where we expected an AP-expansion for the FC-l voids since $\qap^l > 1$, and an AP-contraction for the FC-u void since $\qap^u < 1$.
Note that after correcting for the AP change of volume (right panel), both curves coincide with the z-space VSF of reference remarkably well for all radii of interest.
This is better appreciated by looking at the lower panels, where we show the corresponding fractional differences of void counts $\Delta N_{\rm v }/N_{\rm v } = (N_{\rm v }^{\rm FC}-N_{\rm v }^{\rm zs})/N_{\rm v }^{\rm zs}$ as an indicator of the quality of the correction.
Note that after the correction, the differences between the fiducial and the z-space abundances are reduced to $\Delta N_{\rm v }/N_{\rm v } < 0.2$ in the worst case.
Furthermore, this is also a clear signature that this effect is independent of the other z-space systematics.

We arrive here at the fifth important conclusion of this work: the volume of voids is also affected by the cosmological metric assumed to measure distances, which manifests as an overall expansion or contraction, depending on the chosen fiducial parameters.
Moreover, this effect is independent of any other z-space systematics, and can be statistically quantified as a change of radius by a factor $\qap$.
These results give support to the AP change of volume postulated in Section~\ref{subsec:effects_ap}.

\subsection{Expansion effect correction}
\label{subsec:vsf_RSD}

\begin{figure*}
    \includegraphics[width=\columnwidth]{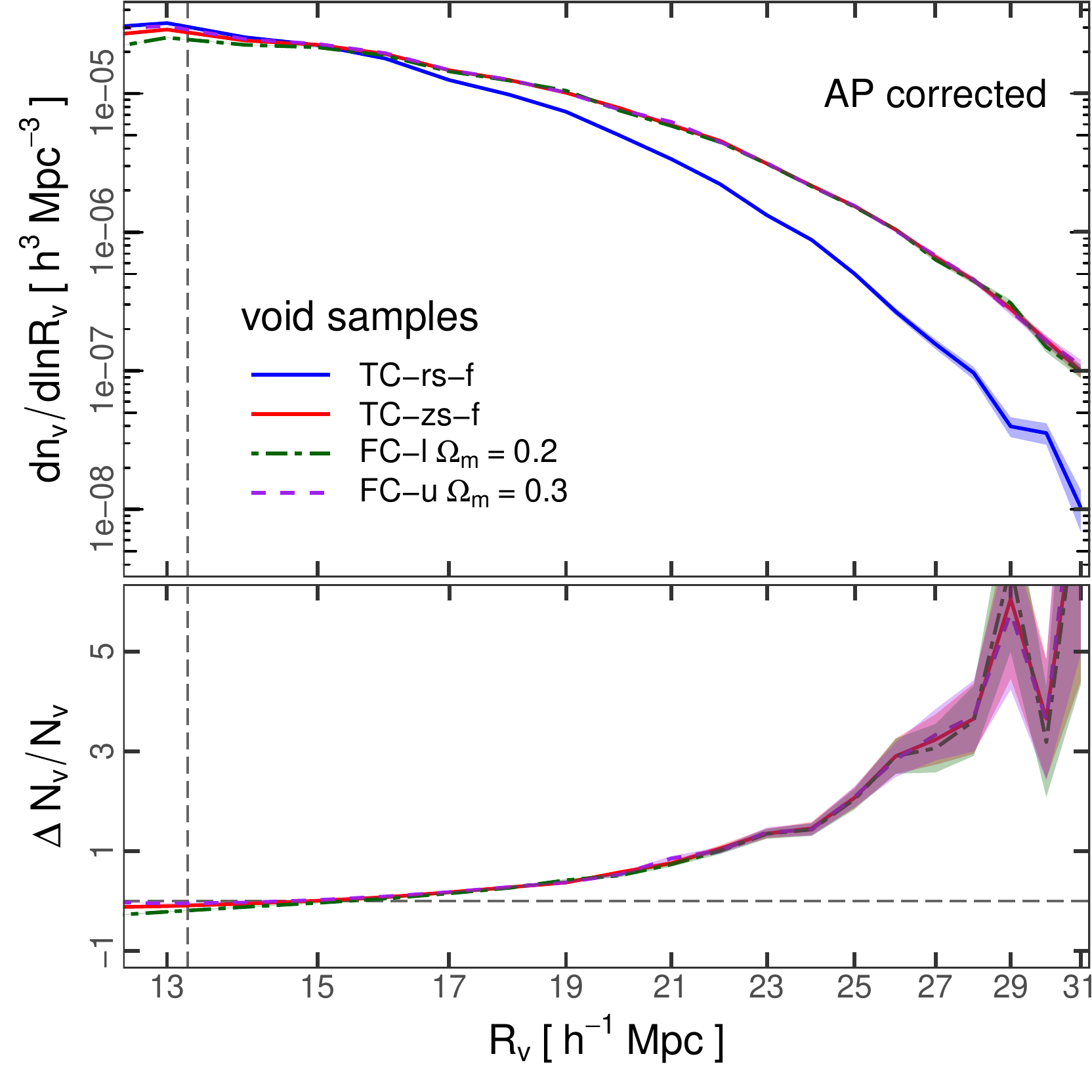}
    \includegraphics[width=\columnwidth]{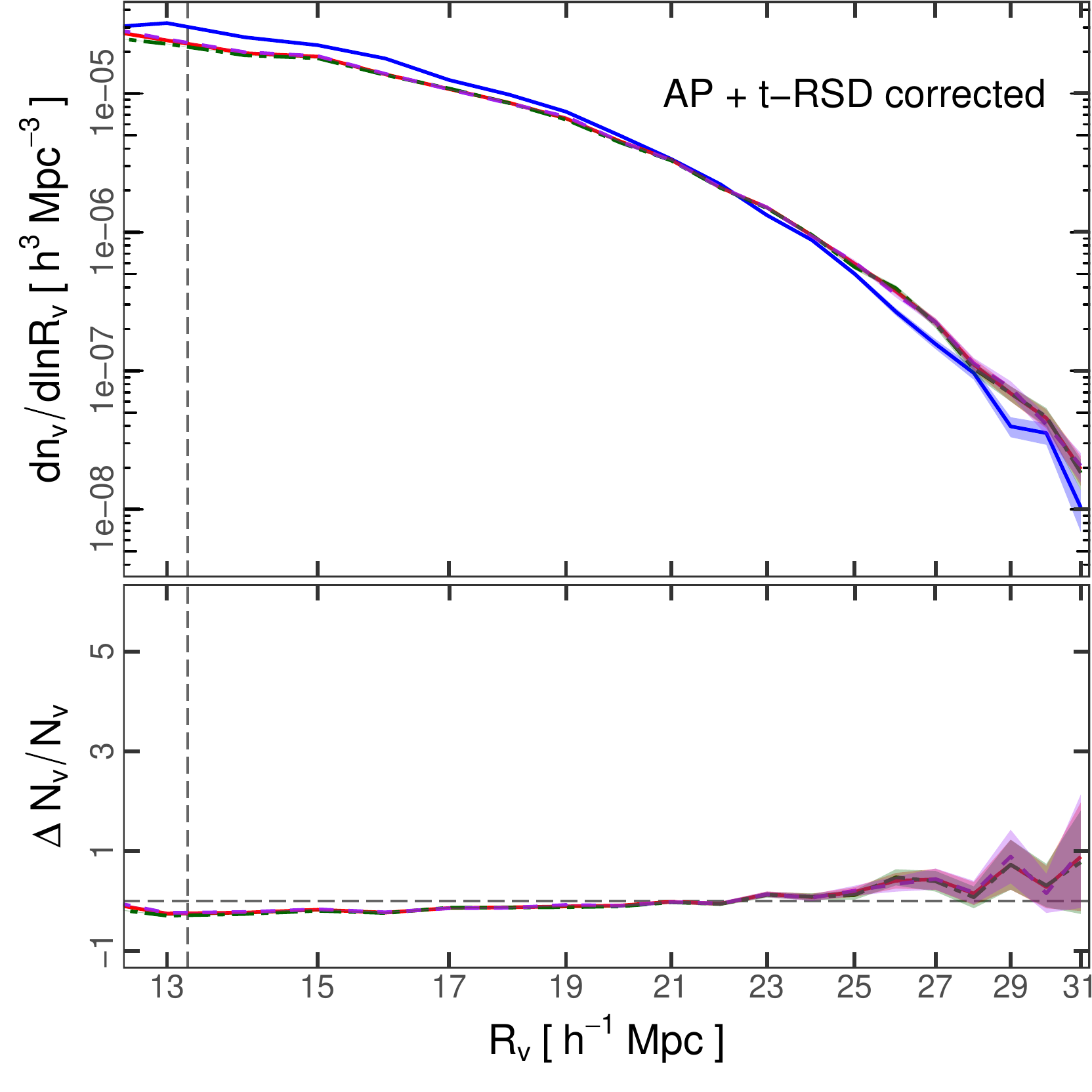}
    \caption{
    t-RSD expansion effect correction of void abundances.
    \textit{Left panel.}
    The same as the right panel of Fig.~\ref{fig:FC_VSF_AP}, except for the fact that the fractional differences were taken with respect to the r-space sample.
    \textit{Right panel.}
    The same as the left panel, but after correcting the abundances for the t-RSD expansion effect.
    }
    \label{fig:FC_VSF_AP_RSD}
\end{figure*}

In this subsection, we discuss the second step of the correction: the t-RSD expansion effect.
The goal now is to recover the r-space VSF of reference.
This is shown in Fig.~\ref{fig:FC_VSF_AP_RSD}.
The left panel is the same as the right panel of Fig.~\ref{fig:FC_VSF_AP}, except for the fact that the fractional differences are referred now to the r-space sample:
$\Delta N_{\rm v }/N_{\rm v } = (N_{\rm v }^{\rm FC}-N_{\rm v }^{\rm rs})/N_{\rm v }^{\rm rs}$.
The right panel shows the correction per se.
This is satisfactory for all radii of interest, although there are some appreciable deviations at the smallest scales.
Note that the large differences between z-space and r-space voids, already noted in Fig.~\ref{fig:TC_VSF}, which can be $\Delta N_{\rm v}/N_{\rm v} > 4$ at the largest scales, have been reduced to $\Delta N_{\rm v}/N_{\rm v} < 0.8$ in the worst case.

\begin{figure}
    \includegraphics[width=\columnwidth]{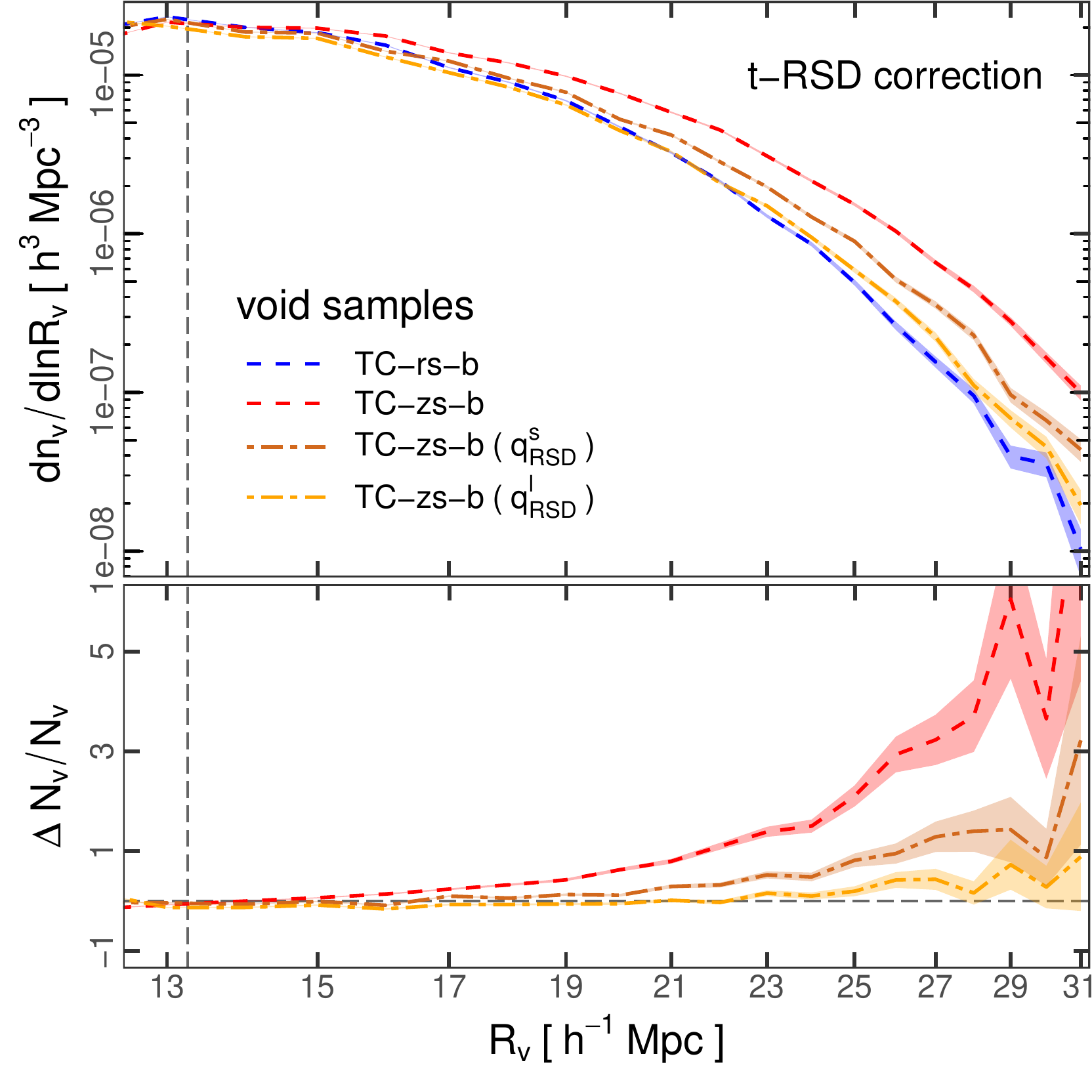}
    \caption{
    t-RSD expansion effect correction of void abundances for the bijective samples taken from the TC catalogues.
    \textit{Upper panel.}
    The red and blue dashed lines are the VSFs of the z-space and r-space samples, the same as in Fig.~\ref{fig:TC_VSF}.
    The brown dot-dashed line represents the correction using Eq.~(\ref{eq:q1_rsd}) with the factor $\qrsd$.
    The orange dot-dashed line represents the correction using Eq.~(\ref{eq:q2_rsd}) with the factor $\qrsdb$.
    \textit{Lower panel.}
    Fractional differences of void counts between the uncorrected and corrected z-space samples with respect to the r-space one, showing the effectiveness of the correction.
    Note that $\qrsdb$ performs better than $\qrsd$.
    A comparison with Fig.~\ref{fig:FC_VSF_AP_RSD} demonstrates that the deviations at small radii are due to the contamination of non-bijective voids at these scales.
    }
    \label{fig:TC_VSF_RSD}
\end{figure}

For the analysis up to here, we have used the full r-space and z-space samples as references (TC-rs-f and TC-zs-f respectively).
This was motivated by the fact that, in the spirit of the bijective mapping analysis, the full and bijective samples can be treated indistinctly.
Moreover, we have used Eq.~(\ref{eq:q2_rsd}) (with $\qrsdb$) to correct for the expansion effect instead of Eq.~(\ref{eq:q1_rsd}) (with $\qrsd$).
In order to test the impact of the impurity of the reference samples regarding the bijective filtering, and the performance of both RSD factors, we repeated the analysis of the last paragraph using now the bijective r-space and z-space samples as references (TC-rs-b and TC-zs-b respectively).
As the AP correction works well at all scales, we have put aside the FC void samples, and only focused on correcting the bijective z-space VSF towards the r-space one.
This is shown in Fig.~\ref{fig:TC_VSF_RSD}.
Note that the z-space VSF (red dashed line) and the r-space VSF (blue dashed line) are the same as in Fig.~\ref{fig:TC_VSF}.
The brown dot-dashed line represents the correction made with $\qrsd$, whereas the orange dot-dashed one, the correction with $\qrsdb$.
Two conclusions can be made.
First, $\qrsdb$ performs better than $\qrsd$, specially at larger scales.
This confirms our suggestion that $\qrsdb$ is more suitable than $\qrsd$ for larger voids, the ones of interest for cosmological analyses.
Second, unlike Fig.~\ref{fig:FC_VSF_AP_RSD}, there are not appreciable deviations at small scales.
Therefore, these deviations can be attributed to the contamination of non-bijective voids in the full samples at these scales.

\subsection{Free of off-centring effect}
\label{subsec:vsf_offcentring}

The achievements of the correction process proves another important fact: the VSF is unaffected by the v-RSD off-centring effect.
This was implicitly assumed in the two-step correction, and constitutes the sixth important conclusion of this work.

In summary, the only two necessary ingredients to correct an observational VSF are the $\qap$ and $\qrsdc$ factors, which relate void radii in r-space and z-space.
As we discussed previously, these factors are only two constants of proportionality, independent of the scale, and strongly cosmology dependent: $\qap$ depends only on the background cosmological parameters, whereas $\qrsdc$ depends only on $\beta$, encoding different cosmological information in a decoupled way.
Therefore, the framework developed in this work must be combined with the excursion set used to model void abundances in order to obtain unbiased cosmological constraints from redshift surveys.
This is the seventh and last important conclusion of this work.


\section{Conclusions}
\label{sec:conclusions}

Cosmic voids are promising cosmological probes provided that the z-space systematics that affect their properties are properly treated.
One approach is to use a reconstruction technique to recover the r-space position of tracers before applying the void finding step.
While this method has proved to be accurate in recovering the r-space void statistics, such as the void size function and the void-galaxy correlation function, and in extracting cosmological information from them, it looses the physical information about the structure and dynamics of voids that manifest when they are identified in z-space.

In this work, we explored an alternative approach: we analysed the void finding method in order to understand physically the underlying z-space systematics.
We used a spherical void finder and made a statistical comparison between the resulting real and redshift-space voids, in the context of the four hypotheses commonly assumed to model RSD around voids, which are only valid for voids identified in r-space, and are violated for those identified in z-space: (1) void number conservation, (2) isotropy of the density field, (3) isotropy of the velocity field, and (4) invariability of centre positions.

The main conclusions of this work can be summarised in the following statements.

\begin{enumerate}

\item[1.]
There is a \textit{bijective mapping} between z-space and r-space voids at scales not dominated by shot noise.
This means that each z-space void has a unique r-space counterpart spanning the same region of space and vice-versa.
In this context, condition (1) of void number conservation is not violated.

\item[2.]
Voids in z-space are systematically bigger than their r-space counterparts.
This can be understood as an \textit{expansion effect} and statistically quantified as an increment in void radius by a constant factor $\qrsd$ (Eq.~\ref{eq:q1_rsd}).
Actually, the slightly modified factor $\qrsdb$ (Eq.~\ref{eq:q2_rsd}) has proved to be more suitable for larger voids, the ones of interest for cosmological studies.
For this analysis, we assumed the validity of hypotheses (2) and (3) concerning the isotropy of the density and velocity fields in r-space in order to explain a z-space phenomenon, even if this isotropy is no longer valid for z-space voids.
This expansion effect is a by-product of the RSD induced by \textit{tracer dynamics} (t-RSD) at scales around the void radius.

\item[3.]
Void centres are systematically shifted along the LOS when they are identified in z-space.
It is a direct consequence of the violation of hypothesis (4) concerning the invariability of centre positions.
This \textit{off-centring effect} can be statistically quantified by means of Eq.~(\ref{eq:void_zspace}).
Hence, it constitutes a different class of RSD induced by large scale flows in the matter distribution.
Interpreting voids as whole entities moving in space with a net velocity, this effect can be thought as a by-product of the RSD induced by \textit{void dynamics} (v-RSD).

\item[4.]
The expansion and off-centring effects are statistically independent, since they manifest in observations as two uncoupled effects.

\item[5.]
The volume of voids is also altered by the fiducial cosmology assumed to transform angular positions and redshifts into distances, which manifests itself as an overall expansion or contraction, depending on the chosen fiducial parameters.
This is the \textit{AP change of volume}.
Moreover, this effect is independent of the other two z-space systematics, and can be statistically quantified as a change of radius by a constant factor $\qap$ (Eq.~\ref{eq:q_ap}).
Therefore, all z-space systematics of this paper can be treated separately.

\item[6.]
The void size function is affected by the t-RSD expansion effect and the AP change of volume, but it is free of the v-RSD off-centring effect.
Therefore, an observational VSF can be corrected in order to recover the true underlying r-space VSF by a simple two-step correction given by Eq.~(\ref{eq:q_ap_rsd}).

\item[7]
The AP and RSD constant factors are strongly cosmology dependent: $\qap$ depends only on the background cosmological parameters, whereas $\qrsdc$ depends only on $\beta$, encoding in this way different cosmological information in a decoupled way.
The former encodes information about the expansion history and geometry of the Universe, whereas the latter, about the growth rate of cosmic structures.
Therefore, the framework developed in this work must be combined with the excursion set theory used to model void abundances in order to obtain unbiased cosmological constraints from redshift surveys.

\end{enumerate}

Although the VSF is unaffected by the v-RSD off-centring effect, this is not the case for the void-galaxy correlation function.
In a follow-up paper (Correa et al. in prep.), we will show that this effect plays a significant role, inducing new distortion patterns in observations.
In the literature, only t-RSD are taken into account when modelling this statistic.
The z-space effects studied in this work must be incorporated in any analysis of RSD around voids in order to successfully exploit these cosmic structures as cosmological probes.
This is particularly important in view of the new generation of spectroscopic surveys, such as BOSS, DESI and Euclid, which will probe our Universe covering a volume and a redshift range without precedents.
Even more, besides its cosmological importance, these z-space systematics encode key information about the structural and dynamical nature of voids.




\section*{Acknowledgements}

This work was partially supported by the Consejo de Investigaciones Cient\'ificas y T\'ecnicas de la Rep\'ublica Argentina (CONICET) and the Secretar\'ia de Ciencia y T\'ecnica de la Universidad Nacional de C\'ordoba (SeCyT).
This project has received financial support from the European Union's Horizon 2020 Research and Innovation programme under the Marie Sklodowska-Curie grant agreement number 734374 - project acronym: LACEGAL.
This research was also partially supported by the Munich Institute for Astro- and Particle Physics (MIAPP) which is funded by the Deutsche Forschungsgemeinschaft (DFG, German Research Foundation) under Germany´s Excellence Strategy – EXC-2094 – 390783311.
CMC acknowledges the hospitality of the Max Planck Institute for Extraterrestrial Physics (MPE), where part of this work has been done.
ANR acknowledges the financial support of the Agencia Nacional de Investigaci\'on Cient\'{\i}fica y T\'enica of Argentina (PICT 2016-1975).
NP was supported by Fondecyt Regular 1191813, and ANID project Basal AFB-170002, CATA.
REA acknowledges the support of the ERC-StG number 716151 (BACCO).
Numerical calculations were performed at the computer clusters from the Centro de C\'omputo de Alto Desempe\~no de la Universidad Nacional de C\'ordoba (CCAD, http://ccad.unc.edu.ar).
Plots were made with the ggplot2 package \citep{ggplot2} of the R software \citep{R}.
CMC would like to specially thank Daniela Taborda for helping with the design of Fig.~\ref{fig:effects} and, fundamentally, for her unconditional support.


\section*{Data Availability}

The data underlying this article will be shared on reasonable request to the corresponding author.




\input{voids_zspace.bbl}







\bsp	
\label{lastpage}
\end{document}